\documentclass[10pt,amsmath,amssymb,onecolumn,eqsecnum,prd,aps,superscriptaddress, nofootinbib]{revtex4-2}
\usepackage{silence}
\usepackage{geometry}
\usepackage{multirow}
\usepackage{booktabs}
\usepackage{tabularx}
\usepackage{adjustbox,caption}
\DeclareUnicodeCharacter{03C3}{-}
\DeclareUnicodeCharacter{03C9}{-}
\usepackage{amsmath,amssymb}
\usepackage{amsthm}
\usepackage{mathtools}
\usepackage{appendix}
\usepackage{comment}
\usepackage{graphicx}
\usepackage{grffile}
\usepackage{mathrsfs}
\usepackage{bm}
\usepackage{url,caption}
\usepackage[table]{xcolor}
\usepackage[colorlinks=true
,urlcolor=DARKBLUE
,anchorcolor=DARKBLUE
,citecolor=DARKBLUE
,filecolor=DARKBLUE
,linkcolor=DARKBLUE
,menucolor=DARKBLUE
,linktocpage=true
,pdfproducer=medialab
,pdfa=true]{hyperref}
\usepackage{xcolor}
\usepackage{soul}
\usepackage{algorithm}
\usepackage{algorithmic}
\usepackage{natbib}
\usepackage{siunitx}
\usepackage{acronym}
\usepackage{xspace}
\usepackage{subfloat}
\usepackage[normalem]{ulem}
\usepackage{amssymb}
\usepackage{amsmath}
\usepackage{graphicx}
\usepackage{dcolumn}
\usepackage{float} 
\usepackage{multirow}
\usepackage{amsfonts,wasysym,epsfig,verbatim,subfigure,bm,mathrsfs,lipsum}
\usepackage{bm}
\usepackage{soul}
\usepackage{pifont}
\usepackage{multirow}
\usepackage{natbib}
\usepackage{ragged2e}

\definecolor{MONZA}{HTML}{CF000F}
\definecolor{DARKBLUE}{HTML}{00008b}
\definecolor{DARKMAGENTA}{HTML}{8b008b}
\definecolor{DARKCYAN}{HTML}{00cfc0}
\definecolor{brightpink}{rgb}{1.0, 0.0, 0.5}

\newcommand{\orcidicon}[1]{\href{https://orcid.org/#1}{\includegraphics[height=\fontcharht\font`\B]{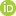}}}

\begin{document}
\title{Impacts of {$f(R, T)$} gravity on neutron stars study within the relativistic mean-field model framework in light of GW170817, Pulsars and NICER data}
\author{Premachand Mahapatra~\orcidicon{0000-0002-3762-8147}}
\email{p20210039@goa.bits-pilani.ac.in}
\author{Prasanta Kumar Das~\orcidicon{0000-0002-2520-7126}}
\email{pdas@goa.bits-pilani.ac.in}
\affiliation{Department of Physics, Birla Institute of Technology and Science-Pilani, K. K. Birla Goa Campus, NH-17B, Zuarinagar, Sancoale, Goa- 403726, India}

\date{\today}

\begin{abstract}
{\color{black}In this work, we investigate the neutron star structure in conservative $f(R, T)$ gravity with $f(R, T)=R+\lambda T$, where $\lambda$ denotes the matter--geometry coupling. The modified stellar structure equations are solved using realistic relativistic mean-field (RMF) equations of state (EOSs), including density-dependent linear models and nonlinear interacting models with meson self-couplings. Theoretical predictions are confronted with multimessenger constraints from heavy pulsars, NICER radius measurements, and GW170817 tidal deformability, imposing $M_{\max}\simeq 2.07\,M_{\odot}$ and $10.62~\mathrm{km}<R_{1.4}<12.83~\mathrm{km}$ to constrain both the EOS parameter space and $\lambda$.

We find that density-dependent EOSs such as DDH$_\delta$ and TW satisfy all observational constraints for specific $\lambda$ ranges, while nonlinear EOSs (NL3, GM1, TM1), despite large maximum masses, fail to simultaneously satisfy radius and tidal bounds even in modified gravity. The maximum neutron star mass is highly sensitive to the matter--geometry coupling and exhibits a strong degeneracy with the EOS, consistent with previous studies. The additional term in the modified Tolman--Oppenheimer--Volkoff equations alters the pressure gradient, affecting EOS stiffness and the speed of sound squared $c_s^2$, while preserving causality ($c_s^2/c^2<1$). Pearson and Kendall analyses reveal a strong negative correlation between mass, radius, and $\lambda$ ($-0.18$ and $-0.23$, respectively). Our results show that modified gravity alone cannot compensate for unrealistic dense-matter physics, highlighting the necessity of realistic EOSs and joint multimessenger constraints, and establish conservative $f(R,T)$ gravity as a viable strong-field extension of General Relativity.

 }

\end{abstract}

\maketitle
\acresetall

\acrodef{SNR}{signal-to-noise ratio}
\acrodef{PSD}{power spectral density}
\acrodef{PBH}{primordial black hole}
\acrodef{NS}{neutron star}
\acrodef{WD}{white dwarf}
\acrodef{GW}{gravitational wave}
\acrodef{NFW}{Navarro--Frenk--White}
\acrodef{LVK}{LIGO--Virgo--KAGRA}

\section{Introduction}

{\color{black}The General Theory of Relativity (GR), formulated by Einstein in 1915, has proven to be remarkably successful in describing gravitational phenomena over a vast range of scales, from laboratory and Solar-System tests to binary pulsars and gravitational-wave observations \cite{Einstein:1915ca, Coley:2016dgx}. Despite these successes, GR faces significant challenges when extrapolated to cosmological and strong-field regimes. In particular, the observed late-time acceleration of the Universe and the formation of large-scale structures cannot be explained within GR without introducing dark energy and dark matter, whose fundamental nature remains unknown \cite{Sahni:2004ai, Khuri:2003hf, Kamionkowski:2007wv}.

Observational evidence for cosmic acceleration has been firmly established through independent probes, including Type Ia supernovae \cite{SupernovaCosmologyProject:1997zqe, SupernovaSearchTeam:1998fmf, Riess:1998dv}, cosmic microwave background radiation anisotropies \cite{Planck:2013pxb, Planck:2015mrs}, baryon acoustic oscillations, weak gravitational lensing, and large-scale structure surveys \cite{Bartelmann:2016dvf, Munshi:2006fn}. Within the standard $\Lambda$CDM cosmological model, these observations are explained by introducing a cosmological constant that accounts for nearly $70\%$ of the total energy density of the Universe, while cold dark matter contributes approximately $25\%$, and the remaining fraction consists of baryonic matter \cite{Weinberg:2013aya}. Although phenomenologically successful, the $\Lambda$CDM model suffers from deep theoretical problems, most notably the cosmological constant problem, which highlights the enormous discrepancy between the observed value of $\Lambda$ and theoretical expectations from quantum field theory \cite{Bull:2015stt, Perivolaropoulos:2021jda}.

These issues have motivated extensive research into extensions of GR, commonly referred to as modified or extended theories of gravity. A comprehensive and systematic overview of such theories, including their cosmological and astrophysical implications, is provided in the seminal review by Nojiri, Odintsov, and Oikonomou \cite{Nojiri:2017ncd}. In this framework, gravity is generalized through higher-order curvature invariants, additional gravitational degrees of freedom, or non-minimal couplings between matter and geometry, offering alternative explanations for cosmic acceleration and strong-field phenomena without invoking unknown dark components \cite{Clifton:2011jh, Nojiri:2010wj}.

From a theoretical perspective, modified gravity theories may be regarded as effective descriptions of gravity beyond GR, potentially arising from quantum corrections, extra-dimensional scenarios, or string-inspired models. Higher-order curvature terms and matter-dependent gravitational couplings naturally emerge in low-energy limits of fundamental theories of gravity \cite{Sotiriou:2008rp, Capozziello:2011et, DeFelice:2010aj}. Consequently, phenomenological frameworks such as $f(R)$ and $f(R,T)$ gravity are not merely ad hoc constructions but provide theoretically motivated extensions capable of capturing deviations from GR in regimes of extreme curvature or matter density.

An important distinction emphasized in modern studies of extended gravity \cite{Nojiri:2017ncd, Clifton:2011jh} is that cosmological and astrophysical tests probe complementary sectors of gravitational dynamics. While cosmology primarily constrains background evolution and linear perturbations, compact objects probe the genuinely nonlinear and strong-field regime, where deviations from GR may be significantly amplified. In particular, relativistic stars and black-hole mimickers provide powerful laboratories for testing gravity beyond GR through observable properties such as mass–radius relations, stability criteria, and tidal responses \cite{Olmo:2019flu, Doneva:2017jop}.

Among modified gravity models, $f(R,T)$ gravity, originally proposed by Harko \textit{et al.} \cite{Harko:2011kv}, extends $f(R)$ gravity by allowing an explicit dependence of the gravitational action on the trace $T$ of the energy–momentum tensor. This feature introduces an effective matter–geometry coupling, which may be interpreted as a phenomenological description of microscopic or quantum matter effects on spacetime geometry \cite{Harko:2014pqa}. Such couplings generically modify the conservation laws and introduce additional force terms, whose physical impact becomes particularly relevant in high-density environments, making neutron stars natural candidates for testing $f(R,T)$ gravity.

Neutron stars are among the densest known objects in the Universe and offer a unique arena for probing the interplay between strong gravity and ultra-dense matter. Their internal structure depends sensitively on both the underlying gravitational theory and the equation of state (EOS) of dense nuclear matter. Consequently, neutron stars provide an excellent testing ground for distinguishing GR from its extensions through observables such as maximum mass, radius, pressure profiles, and tidal deformability \cite{Astashenok:2013vza, Astashenok:2014nua, Feola:2019zqg}.

The emergence of multimessenger astrophysics has further strengthened these tests. The detection of the binary neutron star merger GW170817 by the LIGO–Virgo Collaboration \cite{LIGOScientific:2017vwq} placed the first direct constraints on neutron star tidal deformability, while precise mass measurements of heavy pulsars and recent radius determinations by the NICER mission have significantly narrowed the allowed EOS parameter space. These observations provide unprecedented opportunities to simultaneously constrain dense matter physics and modified gravity theories \cite{Moffat:2020jic, Yagi:2013awa, Berti:2015itd}.

A realistic description of neutron star interiors requires reliable modeling of dense nuclear matter. In this work, we employ relativistic mean-field (RMF) models \cite{PhysRevC.36.2590, PhysRevC.90.055203}, which provide a covariant and microscopically motivated framework for describing strongly interacting matter at supranuclear densities. RMF models naturally incorporate meson-mediated interactions, remain well behaved at high densities, and are particularly suitable for studies of neutron stars in relativistic modified gravity frameworks. Compared to non-relativistic approaches, RMF-based EOSs offer a consistent treatment of relativistic effects and reproduce observed neutron star masses, radii, and tidal deformabilities \cite{PhysRevC.97.045806, PhysRevC.99.045202}.

Several studies have investigated compact stars within $f(R, T)$ gravity using both conservative and non-conservative formulations. Exact interior solutions, anisotropic stellar models, and stability analyses have been explored using simplified EOSs or phenomenological assumptions \cite{Maurya:2019iup, Maurya:2020ebd, Singh:2020iqh, Lobato:2020fxt, Pretel:2021kgl}. More recently, energy-momentum-conserved formulations of $f(R, T)$ gravity have been shown to yield more physically consistent stellar configurations by G.A. Carvalho et al.\cite{Carvalho:2019gzs}. However, a systematic investigation of neutron stars in conserved $f(R, T)$ gravity employing realistic RMF equations of state and confronting the results with up-to-date multimessenger observational constraints remains relatively unexplored.

The primary objective of this work is to fill this gap. We study neutron star structure within an energy-momentum-conserved $f(R,T)=R+\lambda T$ gravity model, employing modern RMF equations of state and confronting the resulting mass–radius relations, pressure profiles, and tidal properties with constraints from GW170817, NICER observations, and massive pulsars. Our analysis provides new insights into the role of matter–geometry coupling in strong-field gravity and highlights the potential of neutron stars as powerful probes of extended gravitational theories.

The paper is organized as follows. In Section~\ref{sec:tov}, we review the basic stellar structure equations. The RMF models employed in this work are presented in Section~\ref{sec:rmfmodel}. The formulation of energy–momentum-conserved $f(R,T)$ gravity is discussed in Section~\ref{sec:frt}, followed by the modified Tolman–Oppenheimer–Volkoff equations in Section~\ref{sec:stellfrt}. Numerical methods and boundary conditions are described in Section~\ref{sec:numbound}. Our results and their astrophysical implications are presented in Section~\ref{sec:resdis}. Finally, we summarize our conclusions in Section~\ref{sec:concl}.}

\section{Neutron Stars in General Relativity} \label{sec:tov}

Neutron Stars, believed to be formed in Type-II supernova explosions, are possible end products of a main-sequence star (“normal” star).  Neutron stars, the most compact stars in the Universe, were given this name because their interior is largely composed of neutrons \cite{1996csnp.book.....G}.

\begin{itemize}
	\item A neutron star is of the typical mass $M \sim 1 - 2 M_\odot$, where $M_\odot \times 10^{33} g$ is the solar mass.
	\item It has the radius of $R \approx  10  -  14~\rm{km} $.
	\item The mass density $\rho$ in such star is roughly 3 times the normal nuclear density (the typical density of a heavy atomic nucleus)  $\rho_{0} \approx  10^{14}~\rm{ g ~cm^{-3}}$ \cite{Ozel:2016oaf, Fraga:2015xha, Sotani:2022ucj, Trautmann:2016ntm}.
\end{itemize}

\subsection*{Tolman–Oppenheimer–Volkoff (TOV) equation}
When considering stellar structure, it is crucial to account for the wide range of densities in neutron stars, from a few $\mathrm{g/cm^3}$ at the surface to more than  $10^{15}~\mathrm{g/cm^3}$ at the core. This variation in density leads to changes in the composition from the centre to the surface, i.e., the EOS varies. According to current theories, a neutron star can be divided into several regions: the atmosphere (characterized by a plasma region with intense magnetic/electric fields), the outer crust, the inner crust, the outer core, and the inner core. To accurately describe these layers, different theories are required: plasma physics and atomic structure theories for the outer regions (outer and inner crust), and many-body theories for high-density strongly interacting systems for the inner and outer cores  \cite{Lattimer:2004pg, Kunz:2022wnj}.

Given these diverse regimes and densities, only the outer crust can be described with high accuracy, and this can be compared with experimental data on atomic nuclei. The EOS for the neutron star interior above nuclear matter density remains unconstrained and is an open question in astrophysics. Nonetheless, there are constraints from microscopic physics such as electric neutrality, beta equilibrium, causality (the speed of sound, $c_s$, must be less than the speed of light, $c$), and the Le Chatelier’s principles ($p \geq 0$ and $dp/d\rho > 0$) that must be satisfied \cite{Alford:2022bpp}.

The uncertainty in describing the neutron star interior results in a wide range of EOS models in the literature, classified as \textit{soft} or \textit{stiff} based on their compressibility and behavior at high densities. These models also differ by matter composition: for the outer core, a $npe\mu$ (neutron-proton-electron-muon) plasma is considered; for the inner core, various possibilities exist, such as fermion/boson condensates, hyperons, pion/kaon condensation, or strange quark matter. The last option, a hybrid neutron star, features hadronic matter surrounding a quark matter core (approximately 3 km depending on the quark matter model) \cite{Alford:2022bpp, Piekarewicz:2022ycz, Nattila:2022evn}.

Using Birkhoff’s theorem\cite{Abbassi:2001ny, Schleich:2009ix} we are free to write the general metric for the stellar interior in the time-independent form, and then we may write :
\begin{align}
	ds^{2} = g_{\mu \nu} dx^\mu dx^\nu =  - e^{2 \psi(r)} dt^{2} + e^{2 \zeta(r)}  dr^{2}  
	+ r^{2} ( d \theta^{2} + \sin^{2} \theta d \phi^{2})
\end{align}
where $g_{0 0} = - e^{2 \psi(r)} , g_{1 1} = e^{2 \zeta(r)} , g_{2 2} = r^{2} , g_{33} = r^{2} \sin^{2} \theta  $ respectively.

\noindent The TOV equations (\textit{i.e} representation of Einstein Equation for the interior of a spherical, static, relativistic compact star(e.g. Neutron star)) are given by 
\begin{equation}
	\frac{dm}{dr}  =  4 \pi r^2 \rho(r)  \label{dmdr}
\end{equation}
\begin{equation}
	\begin{split}
		\frac{dP}{dr} = -  \left(\rho + P \right) \dfrac{m + 4\pi r^3 P}  {r (r - 2m)}  \label{dpdr} \\
	\end{split}
\end{equation}


\noindent Solving the above equations, we get the relationship between Mass($m(r)$) $\&$ Radius(R) and Pressure($P(r)$) $\&$ Energy Density ($\rho$)(where $r$ is the radial coordinate which equals to $R$ at the surface of the star and $R$ is the radial distance.)

\section{Relativistic mean-field models} \label{sec:rmfmodel}
A generic Relativistic mean-field (RMF) models can produce the nuclear matter properties which are relevant to study the neutron star properties. The RMF Lagrangian we use to study the neutron star matter in the present work includes proton, neutron coupled to $\sigma$, $\omega$ and $\vec{\rho}$ mesons and is expressed as follows \cite{Xia:2022dvw}:
\begin{eqnarray}
\mathcal{L}
 &=& \sum_{i=n,p} \bar{\psi}_i
       \left[  i \gamma^\mu \partial_\mu - \gamma^0 \left(g_\omega\omega + g_\rho\rho\tau_i + A q_i\right)- m_i^* \right] \psi_i
\nonumber \\
 &&\mbox{} + \sum_{l=e,\mu} \bar{\psi}_l \left[ i \gamma^\mu \partial_\mu - m_l + e \gamma^0 A \right]\psi_l - \frac{1}{4} A_{\mu\nu}A^{\mu\nu}
\nonumber \\
 &&\mbox{} + \frac{1}{2}\partial_\mu \sigma \partial^\mu \sigma  - \frac{1}{2}m_\sigma^2 \sigma^2
           - \frac{1}{4} \omega_{\mu\nu}\omega^{\mu\nu} + \frac{1}{2}m_\omega^2 \omega^2
\nonumber \\
 &&\mbox{} - \frac{1}{4} {\vec \rho}_{\mu\nu} . {\vec \rho}^{\mu\nu} + \frac{1}{2}m_\rho^2 {\vec \rho}_\mu . {\vec \rho}^\mu  + U(\sigma, \omega),
\label{eq:Lagrange}
\end{eqnarray}
where 

\begin{eqnarray}
\omega_{\mu\nu} = \partial_\mu \omega_\nu - \partial_\nu \omega_\mu, \\
\vec{\rho}_{\mu\nu} = \partial_\mu {\vec \rho}_{\nu} - \partial_\nu {\vec \rho}_{\mu}, \\
A_{\mu\nu} = \partial_\mu A_\nu - \partial_\nu A_\mu
\end{eqnarray}

\noindent with $\tau_n=-\tau_p=1$ being the third component of isospin, $q_i=e (1-\tau_i)/2$ the charge, and $m_{n,p}^* = m_{n,p} + g_{\sigma} \sigma$ the effective nucleon mass. The boson fields $\sigma$, $\omega$, $\rho$, and $A$ acquire mean values with only time components due to time-reversal symmetry. Consequently, the field tensors $\omega_{\mu\nu}$, ${\vec \rho}_{\mu\nu}$(iso-vector field strength tensor), and $A_{\mu\nu}$ are non-zero only for
\begin{equation}
\omega_{i0} = -\omega_{0i} = \partial_i \omega,
 \rho_{i0}  = -\rho_{0i}   = \partial_i  \rho,
  A_{i0}    = -A_{0i}      = \partial_i A.\nonumber
\end{equation}

The nonlinear self-couplings of the mesons are defined as
\begin{equation}
U(\sigma, \omega) = -\frac{1}{3}g_2\sigma^3 - \frac{1}{4}g_3\sigma^4 + \frac{1}{4}c_3\omega^4,  \label{eq:U_NL}
\end{equation}
which effectively represents 
the in-medium effects and are crucial for the covariant density functionals NL3~\cite{PhysRevC.55.540},  TM1~\cite{SUGAHARA1994557}, and GM1~\cite{PhysRevLett.67.2414} used in this work. Alternatively, the in-medium effects can be addressed using density-dependent coupling constants as per the Typel-Wolter approach~\cite{TYPEL1999331}, where
\begin{eqnarray}
g_{\xi}(n_\mathrm{b}) &=& g_{\xi} a_{\xi} \frac{1+b_{\xi}(n_\mathrm{b}/n_0+d_{\xi})^2}
                          {1+c_{\xi}(n_\mathrm{b}/n_0+d_{\xi})^2}, \label{eq:ddcp_TW} \\
g_{\rho}(n_\mathrm{b}) &=& g_{\rho} \exp{\left[-a_\rho(n_\mathrm{b}/n_0 + b_\rho)\right]}. \label{eq:ddcp_rho}
\end{eqnarray}
Here, $\xi=\sigma, \omega$, and $n_\mathrm{b} = n_p+n_n$ is the baryon number density with $n_0$ as the saturation density. Alongside the nonlinear models, we have also incorporated the density-dependent covariant density functionals  TW~\cite{TYPEL1999331}, DDH$_\delta$ ~\cite{GAITANOS200424}, and  DD2~\cite{PhysRevC.81.015803},  where the nonlinear self-couplings in Eq.~(\ref{eq:U_NL}) vanish with $g_2=g_3=c_3=0$. For completeness, the parameter sets utilized in this study with $a_{\sigma, \omega}=1$ and $b_{\sigma, \omega}=c_{\sigma, \omega}=a_\rho=0$ if nonlinear self-couplings are employed.




The total energy of the system is then given by:
\begin{equation}
E=\int \langle {\cal{T}}_{00} \rangle \, \text{d}^3 r, \label{eq:energy}
\end{equation}
with the energy-momentum tensor given by:

\begin{eqnarray}
\langle {\cal{T}}_{00} \rangle
&=& \sum_{i=n,p,e,\mu} \frac {{m^*_i}^4}{8\pi^{2}} \left[x_i(2x_i^2+1)\sqrt{x_i^2+1}-\mathrm{arcsh}(x_i) \right] \nonumber \\
&&   + \frac{1}{2}(\nabla \sigma)^2 + \frac{1}{2}m_\sigma^2 \sigma^2 + \frac{1}{2}(\nabla \omega)^2 + \frac{1}{2}m_\omega^2 \omega^2 + c_3\omega^4 \nonumber \\
&&   + \frac{1}{2}(\nabla \rho)^2 + \frac{1}{2}m_\rho^2 \rho^2
     + \frac{1}{2}(\nabla A)^2 - U(\sigma, \omega),
\label{eq:ener_dens}
\end{eqnarray}
where $x_i\equiv \nu_i/m^*_i$ with $m_e^*=m_e = 0.511$ MeV and $m_\mu^*=m_\mu = 105.66$ MeV. \\

Working within the Thomas-Fermi approximation (TFA), one finds the optimal density distributions $n_i(\vec{r})$ by minimizing the total energy $E$ under the constraints of given total particle numbers $N_i=\int n_i \text{d}^3 r$, dimension $D$, and Wigner-Seitz (WS) cell size $R_\mathrm{W}$, while ensuring that the chemical potentials remain constant:

\begin{equation}
\mu_i(\vec{r}) = \sqrt{{\nu_i}^2+{m_i^*}^2} + \Sigma^\mathrm{R} + g_{\omega} \omega + g_{\rho}\tau_{i} \rho + q_i  A = \text{constant}. \label{eq:chem_cons}
\end{equation}

It is important to account for the "rearrangement" term $\Sigma^\mathrm{R}$ if density-dependent couplings are included in the Lagrangian density~\cite{LENSKE1995355}:
\begin{equation}
\Sigma^\mathrm{R}=
 \frac{\text{d} g_\sigma}{\text{d} n_\mathrm{b}} \sigma n_\mathrm{s}+
   \frac{\text{d} g_\omega}{\text{d} n_\mathrm{b}} \omega n_\mathrm{b}+
   \frac{\text{d} g_\rho}{\text{d} n_\mathrm{b}} \rho \sum_i\tau_i n_i.
\label{eq:re_B}
\end{equation}

The fundamental relation gives the pressure ;
\begin{align}
    P = \sum_{i} \mu_{i} n_{i} - E + n_{B}  \Sigma^{\mathrm{R}}
\end{align}

So, for our work, we have chosen unified Hardonic EOS ($P$(pressure) and $E$(energy)) with different parametrization viz:
\begin{enumerate}
    \item The nonlinear interacting models with self-coupling NL3, GM1, TM1:``The well-known NL3 model incorporates non-linear interactions by including only the self-coupling term of the $\sigma$-meson while omitting the cross-coupling terms."
    \item The density-dependent linear models DD2, DDH$_\delta$, TW: ``The density-dependent interaction, based on the experimental values of the proton and neutron masses, \( m_p \) and \( m_n \), enables this model to accurately describe the composition and thermodynamic properties across a wide range of densities."
\end{enumerate}
\vspace*{0.05in}

\section{Field Equations in {$f(R,T)$} gravity} \label{sec:frt}
The $f(R,T)$ theories of gravity \cite{Harko:2011kv,Tretyakov:2018yph,Ashmita:2022swc,Arora:2020xbn,Deb:2022syd} is a generalization of $f(R)$ theories of gravity \cite{Sotiriou:2008rp,DeFelice:2010aj,Nojiri_2017} given by Harko and Lobo describing the couplings of the trace of the energy-momentum tensor with the curvature of space-time. The action in $f(R, T)$ theories of gravity, depends on a general function of the Ricci scalar $R$ and the trace of the energy-momentum tensor $T$ and is given by \cite{Harko:2011kv, Carvalho:2017pgk} :

\begin{flalign} 
	S = \int d^{4} x \sqrt{- g} \left[     \dfrac{1}{16 \pi} f(R,T) + \mathcal{L}_{m} \right]                           \label{action}
\end{flalign}

\noindent where  $\mathcal{L}_{m}$ is the matter Lagrangian density and  $g$ is the determinant of the metric tensor $g_{\mu \nu}$ and $G$ is the Newtonian constant of Gravitation. In this work, we use the metric signature $(-, +, +, +)$ and $G = 1 = c$ unit .




Varying the action Eq. \eqref{action} with respect to the metric $g_{\mu \nu}$, we get the modified gravity field equations as,
\begin{gather}
	f_{R} (R, T) R_{\mu \nu} - \dfrac{1}{2} f(R, T) g_{\mu \nu}   +   
	(g_{\mu \nu } \Box - \nabla_{\mu} \nabla_{\nu})f_{R} (R, T)  \nonumber \\ 
	=  8 \pi T_{\mu \nu} - f_{T} (R, T) (T_{\mu \nu} + \Theta_{\mu \nu}),     \label{frt}
\end{gather}
\noindent where  $f_{R}(R, T) = {\partial f(R, T) }/{\partial R}$, $f_{T} (R, T) = {\partial f(R, T) }/{\partial T}$ and $\Box (= \nabla_\mu \nabla^\mu)$  is the d'Alembertian Operator with $\nabla_{\mu}$  representing the covariant derivative. $T_{\mu\nu}$ is the energy-momentum tensor, which is given by
\begin{align*}
 T_{\mu \nu} =  g_{\mu \nu} \mathcal{L}_{m} - 2 \dfrac{\delta \mathcal{L}_{m}}{\delta g^{\mu \nu}} 
\end{align*}

Here, $T(=g_{\mu \nu} T^{\mu \nu})$ is the trace of the energy-momentum tensor, and the tensor $\Theta_{\mu \nu}$ is the metric variation of the energy-momentum tensor, which is defined as, 
\begin{align}
	\Theta_{\mu \nu} = g^{\alpha \beta} \dfrac{\partial T_{\alpha \beta}}{ \partial g^{\mu \nu }} = - 2 T_{\mu \nu} + g_{\mu \nu} \mathcal{L}_{m}  - 2 g^{\alpha \beta }  \dfrac{\partial ^{2}    \mathcal{L}_{m}}{\partial g^{\mu \nu}   \partial g^{\alpha \beta}}  \label{theta}
\end{align}
%
The four-divergence of the energy-momentum tensor can be written as  \cite{Carvalho:2017pgk, Moraes:2018uji,Pretel:2021kgl,Lobato:2020fxt}
\begin{align}
	\nabla^{\mu} T _ {\mu \nu}  = \dfrac{f_{T} (R, T)}{ 8 \pi - f_{T} (R, T)} \times \nonumber \\
	\left[ (T_{\mu \nu} + \Theta_{\mu \nu}) \nabla^{\mu} \ln f_{T}   +\nabla^{\mu} \Theta_{\mu \nu} - \dfrac{1}{2} g_{\mu \nu} \nabla^{\mu} T \right]       \label{nabla}
\end{align}



To describe the matter source of stellar structure, we chose the energy-momentum tensor of a perfect fluid, such that
\begin{align}
	T_{\mu \nu} = \left(\rho + P \right) u_{\mu} u_{\nu}  - P g_{\mu \nu},  \label{tmunu}
\end{align}
where $\rho$ and $P$ represent the energy density and the pressure of the fluid, $u^{\mu}(=(1,0,0,0))$ is the time-like fluid four-velocity(comoving) vector with $u_\mu u^\mu= -1$.  

 According to Mendoza \textit{et al.} \cite{Mendoza:2020bzc} the matter Lagrangian density for an ideal fluid  is $\mathcal{L}_m = - \rho $ and Faraoni claimed that \cite{Faraoni:2009rk}, ``These Lagrangians ($\mathcal{L}_m = -\rho $ or , $\mathcal{L}_m = -P$) are equivalent when the fluid couples minimally to gravity and not otherwise. In the presence of nonminimal coupling they give rise to two distinct theories with different predictions."  Here we take $\mathcal{L}_m = - \rho$ and this gives $\Theta_{\mu \nu} =  - 2 T_{\mu \nu} - \rho  g_{\mu \nu}$. 


\noindent Accordingly, the field equations \eqref{frt} lead to 

\begin{align}
	f_{R} R_{\mu \nu} - \dfrac{1}{2} f g_{\mu \nu}   +   
	(g_{\mu \nu } \Box - \nabla_{\mu} \nabla_{\nu})f_{R}  \nonumber \\ 
	= 8 \pi T_{\mu \nu} + f_{T} (T_{\mu \nu} + \rho  g_{\mu \nu})          \label{modfrt}
\end{align}

\noindent We consider a particular case of $f(R, T)$ theory where $f(R, T) = R + h(T)$  \cite{Pretel:2021kgl}. 

With $f_{R} = 1$ and $f_{T} = h_{T} = \frac{\partial h(T)}{\partial T}$, the eqs. \eqref{modfrt} and \eqref{nabla}  take the following form, respectively,
%
\begin{flalign}
	G_{\mu \nu } = 8 \pi T_{\mu \nu} + \dfrac{1}{2} h g_{\mu \nu}  + (T_{\mu \nu} + \rho  g_{\mu \nu})   h_{T} \label{gmunu}
\end{flalign}
\begin{flalign}
	\nabla^{\mu} T _ {\mu \nu} = - \dfrac{h_{T}}{ 8 \pi + h_{T}} \times \nonumber \\
	\left[ (T_{\mu \nu} + \rho  g_{\mu \nu}) \nabla^{\mu} \ln h_{T}   +\nabla_{\nu} \left( \rho  + \dfrac{1}{2} T    \right) \right] \label{conservT}
\end{flalign}
%
%
\noindent where $G_{\mu \nu}~(= R_{\mu\nu} - \frac{1}{2} g_{\mu\nu} R)$ is the Einstein tensor. If $T_{\mu\nu}$ is a conserved quantity \footnote{In an astrophysical level, say, in the construction of the
Tolman-Oppenheimer-Volkoff (TOV) equation, $\nabla^{\mu} T_{\mu \nu}  \neq 0$ cannot be interpreted in the same way as in cosmology. Although the creation of matter in the universal scale is a consequence of processes occurring in a quantum scale, the creation or destruction
of matter particles shall not occur in a static analysis, such
as the TOV equation \cite{dosSantos:2018nmu}}, then $\nabla^{\mu} T_{\mu \nu} =0$ gives (following eqn. (\ref{conservT}))
\begin{flalign}
	(T_{\mu \nu} + \rho  g_{\mu \nu} )\nabla^{\mu} \ln h_{T} + \nabla_{\nu} \left( \rho  + \dfrac{1}{2} T    \right) = 0 \label{conservTmunu}
\end{flalign}

\subsection{Case: $f(R, T)   = R + h(T) $ }

In this work, we adopt a phenomenological approach where we assume the functional form $h(T) = \lambda T$, motivated by its simplicity and widespread use in the literature. We then explore its implications across a range of realistic RMF-based EOSs.Considering $h(T) = \lambda T$, where $\lambda$ is the modified gravity parameter, usually taken to be constant values :
\begin{equation*}
    h_T = \lambda, \quad h_{TT} = 0
\end{equation*}

\subsubsection{Conservation Condition:}

Substituting $h_T = \lambda$ into the conservation equation (\ref{conservTmunu}), which gives:
\begin{align}
    (T_{\mu\nu} + \rho g_{\mu\nu}) \nabla^\mu \ln \lambda + \nabla_\nu \left(\rho + \frac{1}{2} T \right) = 0
\end{align}

Since $\ln \lambda $ is constant, $\nabla^\mu \ln \lambda = 0$. The equation reduces to:
\begin{equation}
    \nabla_\nu \left(\rho + \frac{1}{2} T \right) = 0
\end{equation}

Using $T = -\rho + 3P =$ trace of the energy-momentum tensor $T_{\mu \nu}$ for a perfect fluid, we get:
\begin{equation*}
    \rho + \frac{1}{2} T = \rho + \frac{1}{2} (-\rho + 3P) = \frac{\rho}{2} + \frac{3P}{2}
\end{equation*}

Thus:
\begin{equation}
    \nabla_\nu \left(\frac{\rho}{2} + \frac{3P}{2}\right) = 0 \label{condcons}
\end{equation}

Taking $\nu =1$ in equation(\ref{condcons}), we have;
\begin{equation}
    \frac{1}{2} \partial_{r} (\rho + 3 P) = 0  \quad \Rightarrow \frac{\partial}{\partial r} (\rho + 3 P) =0 \label{conslam}
\end{equation}

As $\partial_{r} = \partial / \partial r$ reflecting the radial derivative. So, eqn. (\ref{conslam})
reflecting conserved dynamics with a specific balance between energy density ($\rho$) and pressure($P$) which is satisfied as long as $\partial_r \rho$ and $\partial_r P$ are consistent with the EOS.There are no additional constraints on the form of the EOS or on $T$ because $h_T = \lambda$ is constant and does not introduce nonlinear dependencies. Therefore, \textbf{$f(R, T) = R + \lambda T$ satisfies the conservation equation.}

As dos Santos et. al \cite{dosSantos:2018nmu} \&  F.G. Alvarenga et al. \cite{Alvarenga:2013syu} already mentioned $\nabla^{\mu} T _ {\mu \nu}=0$ under certain conditions,  we found $R+ \lambda T $ satisfies this condition whether any deviation from it,  which doesn't meet the same. So, we will see in our analysis how this model affect the stellar structure equation.

\section{Equation of Stellar Structure in {$f(R, T)$} gravity} \label{sec:stellfrt}
\subsection{ Hydrostatic equilibrium - Modified TOV equation}

To obtain the Tolman-Oppenheimer-Volkoff equations for a static(non-rotating), relativistic compact Neutron star(NS) in the minimally coupled $f(R, T)$ model, we adopt the general spherically symmetric metric as
\begin{equation}\label{13}
ds^2 = -e^{2\psi}dt^2 + e^{2\zeta}dr^2 + r^2(d\theta^2 + \sin^2\theta d\phi^2) ,
\end{equation}
where $x^\mu = (t, r, \theta, \phi)$ are the Schwarzschild  coordinates, and 
$g_{0 0} = - e^{2 \psi} , g_{1 1} = e^{2 \zeta} , g_{2 2} = r^{2} , g_{33} = r^{2} \sin^{2} \theta  $ respectively.  
 
 Now for a star in a state of hydrostatic equilibrium, the metric components and the thermodynamic quantities like density and pressure do not depend on $t$, only depend on radial coordinate $r$. This leads to the metric potentials $\psi = \psi(r)$ and $\zeta = \zeta(r)$ and the components of the energy-momentum tensor $T^{0}_{0} = -\rho_{0}(r)$ and $T^{1}_{1}= T^{2}_{2}= T^{3}_{3} = P_{0}(r) $, where $\rho_0,~P_0$ (denoted by a lower index $0$) correspond to their equilibrium values.

Now taking the form of $f(R, T)$ of the minimally coupled modified gravity theory as 
%
$$f(R, T) = R + h(T)$$ where $h(T) =  \lambda T  $, $f_{R} =1$ and $f_{T} = h_{T} =  \lambda  $ for $R + \lambda T $ model. 

The TOV equations for the interior of a spherical symmetric, non-rotating relativistic Neutron star(NS) in this modified gravity theory can be derived from the modified Einstein equations (eqn. \ref{gmunu}) and are as follows: 
%
\begin{align}
\frac{dm(r)}{dr} &= 4\pi r^2\rho - \frac{r^2}{4}h(T),   \label{dmdrMod}
\end{align}
\begin{equation}
	\begin{split}
   \frac{dP(r)}{dr} = &-(\rho + P)\bigg[ \frac{m}{r^2} + 4\pi r P \\
  &+ \frac{r}{2}\left( \frac{h(T)}{2} + (\rho+ P)h_T \right) \bigg] \left( 1- \frac{2m}{r} \right)^{-1},  
  \end{split}  \label{dpdrMod} 
%
\end{equation}



\noindent The total mass of the compact star is given by $M = m(r_{sur})$ where $r_{sur}$ denotes the radial coordinate at the stellar surface where the pressure vanishes, i.e. $P(r = r_{sur}) = 0$.  It is evident that when $h(T) = 0$ one recovers the equation. (\ref{dmdr}) and eqn. (\ref{dpdr}), The traditional TOV equations in the pure GR case.

\section{Neutron stars in {$f(R, T)$} gravity}\label{sec:numbound}

\subsection{About the Numerical Method}
We solve the stellar structure equations \eqref{dmdrMod} and \eqref{dpdrMod} numerically using the 4th-order Runge-Kutta method for different values of central densities $\rho_c$($\sim 10^{18}  kg /m^{3}$), and different modified gravity parameters $\lambda$.  \\
We use the  boundary conditions in $f(R,T)$ gravity theory  at the centre $(r = 0)$ 
\begin{equation}
	m(0) =0 , ~ \rho(0) = \rho_{c} ~{\rm and}~ P(0) = P_c
\end{equation}
which is the same as in the usual GR theory (from this section we take  $\rho$ as the energy density of the neutron star.) At the surface of the star $r = R$, the pressure vanishes, i.e., $P(r=R) = 0$.

The metrics of the interior line element and the exterior line element are smoothly connected at the surface $r=R$ by $e^{2 \psi(r)} = e^{- 2 \zeta(r)} = 1 - \frac{2 M}{R}$ where $M$ corresponds to the total stellar mass.
\section{Results and  discussions}  \label{sec:resdis}

\begin{widetext}

\begin{figure}[H]
\centering
\subfigure[\ Combined Equation of State Plots ]{\includegraphics[width=0.49\linewidth]{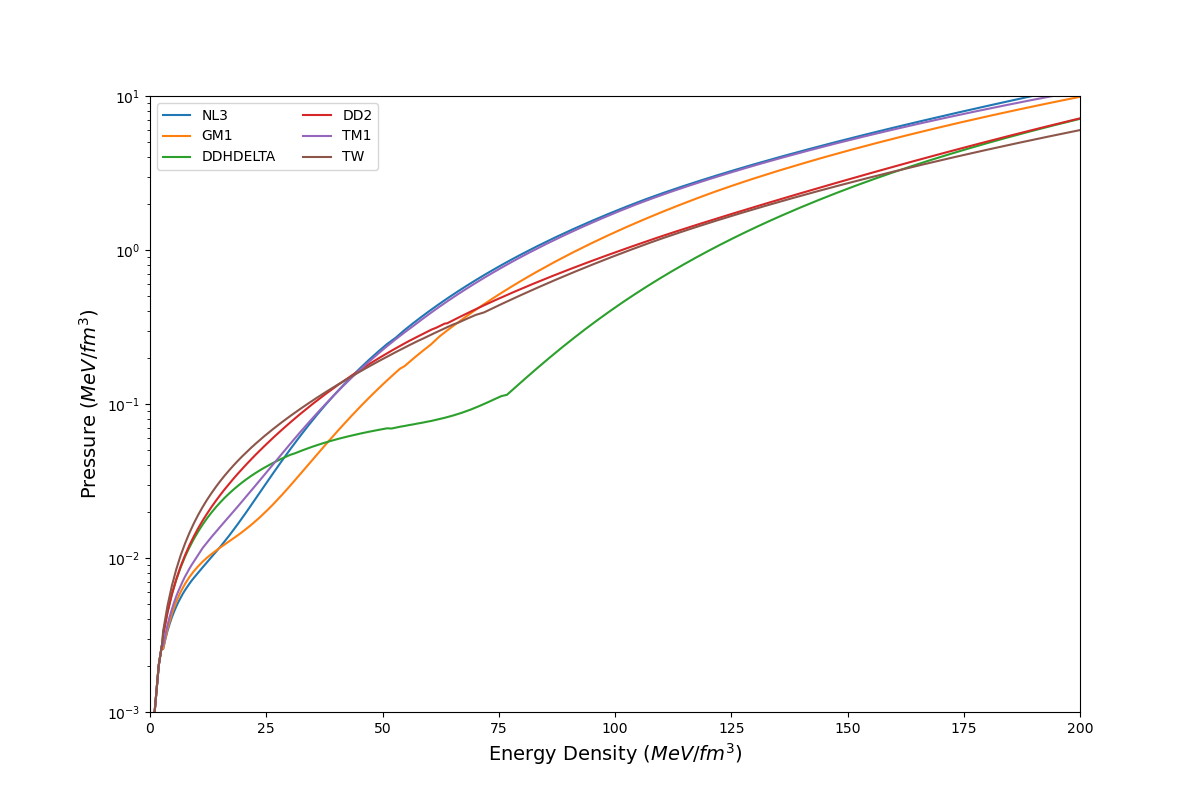}\label{fig:EOS}} %
\subfigure[\  Speed of Sound Squared vs Number Density (in units of $c^2$) ]{\includegraphics[width=0.49\linewidth]{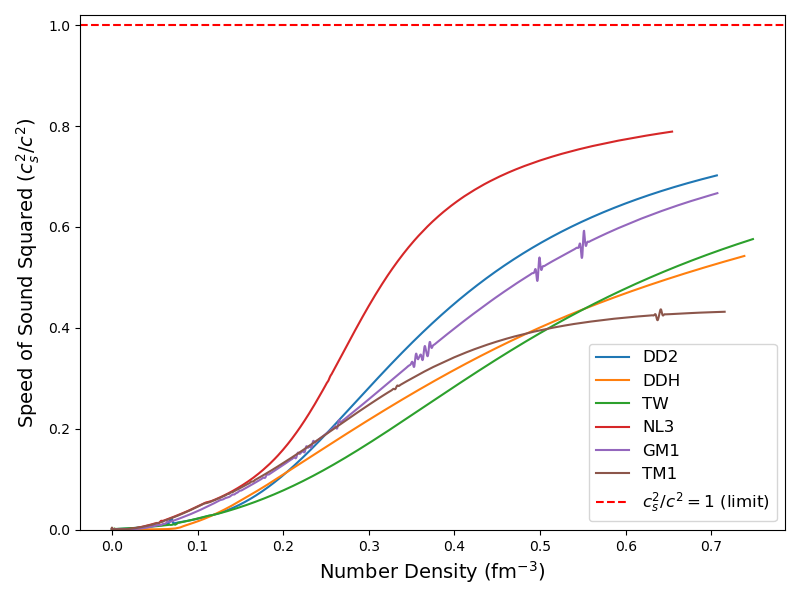}\label{fig:speed of sound}}
\caption{\justifying Pressure $P(\rho)$ (in $MeV/fm^3$) vs Energy Density $\rho$ (in $MeV/fm^3$) and Speed of Sound Squared ($c_s^2 / c^2$) vs Number Density ($fm^{-3}$) for all of the given RMF model EOS} 
\label{alleos-cs2}
\end{figure}

\end{widetext}

Taking RMF-based realistic EOS we study the physical observables of neutron stars. Here we take 6 RMF model Equation of state viz. DD$2$, DDH$_\delta$, TW, NL$3$, GM$1$, and TM$1$  EOS. (The former three models are density-dependent linear EOS and later three are nonlinear interacting EOS). As mentioned in fig \ref{fig:EOS}, we plot the $P(\rho)$ Pressure w.r.t the Energy density ($\rho$);(both of them are in $MeV /fm^3$ units).

We also calculate the speed of sound squared in terms of $c^2$ (where $c$ = speed of light) vs number/ baryonic density (in $fm^{-3}$) which shouldn't cross the maximum speed limit $c_{s}^{2} / c^2 = 1$ i.e the causality condition (the speed of sound, $c_s$, must be less than the speed of light, $c$) for all the given EOS shown in fig. \ref{fig:speed of sound}. The maximum speed of sound squared $c_{s}^{2}$ for  DD$2$, DDH$_\delta$, TW, NL$3$, GM$1$, and TM$1$  EOS are $0.70, 0.54, 0.57, 0.78, 0.66,$ and $0.43$ (in $c^2$ unit) respectively.

\begin{widetext}

\begin{figure}[H]
\centering
\subfigure[\ Mass$(M_\odot)$  vs Radius $R$ (in km) ]{\includegraphics[width=0.49\linewidth]{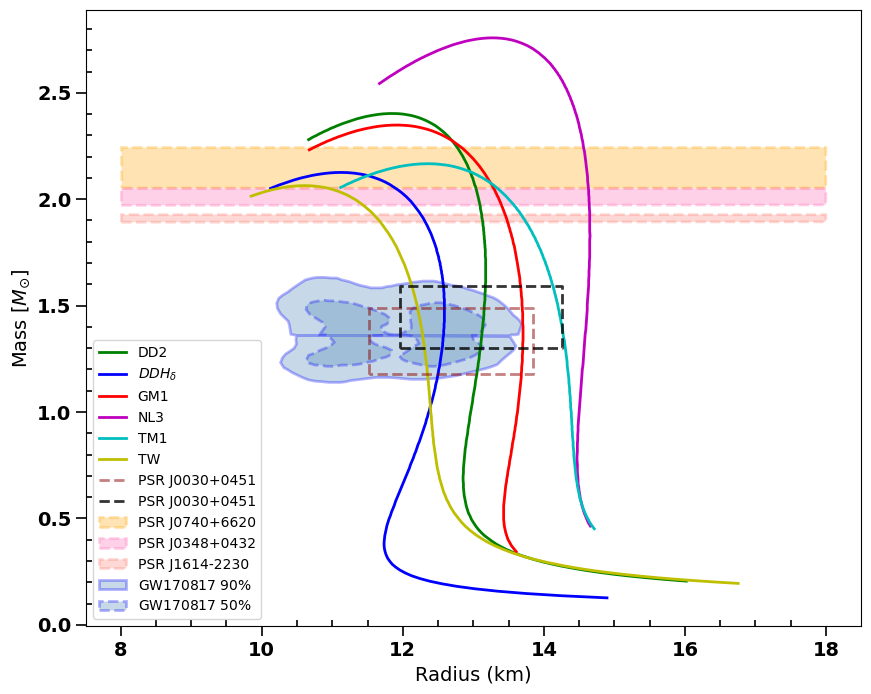}\label{fig:gr_MVRnumberofstars}} 
\subfigure[\  Mass$(M_\odot)$  vs $\rho_c$ ($10^{15}  g /cm^3$ ) ]{\includegraphics[width=0.49\linewidth]{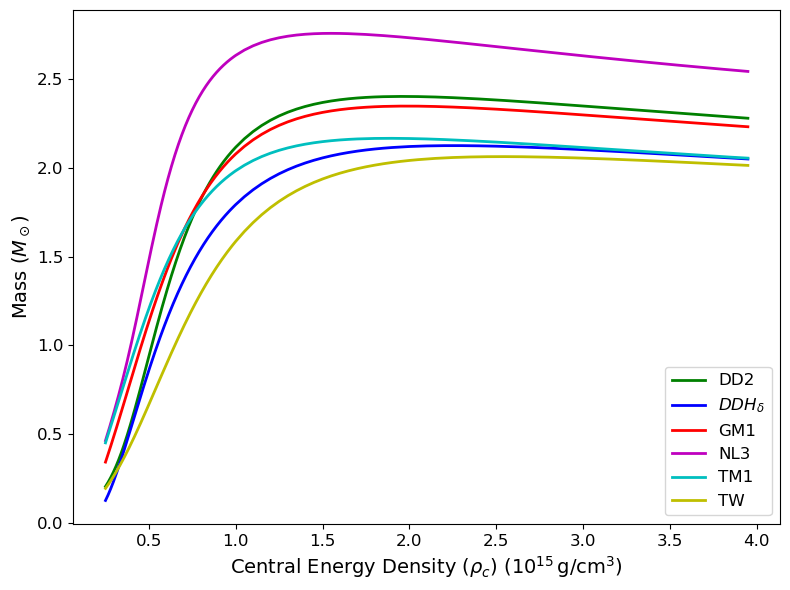}\label{fig:gr_MVRhocnumberofstars}}

\caption{\justifying Mass($M_\odot$) vs Radius ($R$)  and Mass($M_\odot$) vs central energy density $\rho_c$ ($10^{15}  g /cm^3$ ) profile for a number of Neutron Stars in the given range of $\rho_c$ for all $6$ \textbf{RMF EOS}} 
\label{ALL_GR}
\end{figure}

\end{widetext}

While solving the TOV equations of neutron stars for a range of central energy density $\rho_c  = 2.5 \times 10^{14} g /cm^3$ to $4 \times 10^{15} g /cm^3$, we got the one-to-one correspondence plot i.e Mass($M_\odot$) vs Radius(in km) for all EOS in GR as shown in fig (\ref{fig:gr_MVRnumberofstars}) and how Mass($M_\odot$) vary with the range of central energy density $\rho_c$ of number of neutron stars which is shown in fig (\ref{fig:gr_MVRhocnumberofstars}) describing for initial lower values of $\rho_c$ the curve increasing up to the maximum mass point then it is becoming plateau, which should always satisfy the necessary stability criterion $\partial M / \partial \rho_c > 0$.

\textbf{\textit{Observational Constraints:}}

To assess the validity of our findings in fig (\ref{fig:gr_MVRnumberofstars}), we compare them with recent observational constraints. These constraints encompass the GW170817 event, involving the merger of two neutron stars with component masses between $1.17 M\odot$ and $1.6 M_\odot$ having a radius range of $10.62$ km $< R_{1.4} < 12.83$ km i.e the outer and inner grey shaded regions indicate the 90\% (solid) and 50\% (dashed) confidence interval of the LIGO-Virgo analysis for BNS component from the GW170817 event \cite{Nathanail:2021tay, Shibata:2019ctb, abbott2018gw170817, GWOSC_softx}. The rectangular regions enclosed by dotted lines indicate the constraints from the millisecond pulsar PSR J0030+0451 (black \& dark red) NICER x-ray data, which have two independent mass \& radius measurements: $R = 13.02 ^{+ 1.24}_{- 1.06}$ km for mass $M = 1.44 ^{+ 0.15}_{- 0.14} M_{\odot}$ by Miller \textit{etal.} \cite{Miller:2019cac} and radius of $R = 12.71 ^{+ 1.14}_{- 1.19}$ km for mass $M = 1.34 ^{+ 0.15}_{- 0.16} M_{\odot}$ by Riley \textit{etal.} \cite{Riley:2019yda}. Other shaded regions are the mass-radius data of pulsars PSRJ1614--2230 (salmon) (It's a Millisecond Pulsar detected by NICER in 2018) \cite{Takisa:2014sva, Demorest:2010bx, NANOGrav:2017wvv} of mass ($M = 1.908 \smash{\scriptstyle\pm} 0.016 M_{\odot}$) $\&$ radius($ R = 10.30 \sim 9.67$ km) and PSR J0348+0432 (hot pink) detected by Radio Telescope \cite{Antoniadis:2013pzd, Zhao:2016rfv} in 2013 of mass ($M = 2.01 \smash{\scriptstyle\pm} 0.04 M_\odot$) $\&$ radius ($ R= 12.957 \sim 12.246$ km) (noting that the radius values mentioned here are suggested estimates from various authors rather than direct measurements) and recently measurements for PSR J0740+6620 (orange) have reported a radius of $R = 13.7 ^{+ 2.6}_{- 1.5}$ km for mass $M = 2.08 \smash{\scriptstyle\pm} 0.07 M_{\odot}$ by Miller \textit{etal.} \cite{Miller:2021qha} and radius of $R = 12.39 ^{+ 1.30}_{- 0.98}$ km for mass $M = 2.072 ^{+ 0.067}_{- 0.066} M_{\odot}$ by Riley \textit{etal.} \cite{Riley:2021pdl}.

It is worth noting that, some of the density-dependent linear models DD2, DDH$_\delta$, and TW EOS precisely pass the GW170817 regions along with touching the maximum mass values of given Pulsars data. In contrast, the nonlinear interacting models with self-coupling NL3, GM1, TM1 EOS, even crossing the current observational constraints whether from pulsars or NICER observations, are still outside the GW region. By implementation of an extra term $T$ in Einstein's action, we are going to explore how the modified TOV equations \eqref{dmdrMod} \& \eqref{dpdrMod} with these given RMF models try to fit with the GW and NICER mass, radius data.

\subsection{{$R + \lambda T$} model}
\begin{widetext}

\begin{figure}[H]
\centering
\subfigure[\ m$(M_\odot)$  vs $r$ (in km) of DD2 EOS]{\includegraphics[width=0.49\linewidth]{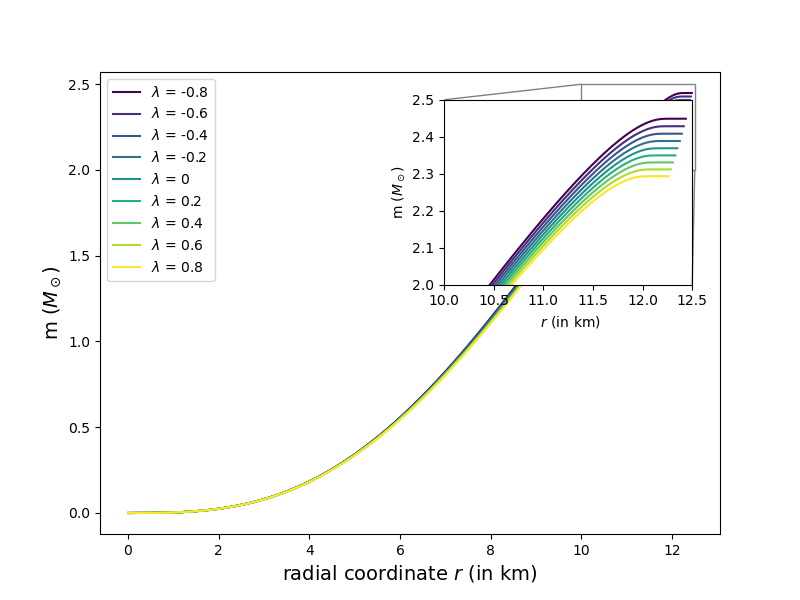}\label{fig:DD2_MVRsingle}} 
\subfigure[\ Pressure(Pa)  vs $r$ (in km) of DD2 EOS]{\includegraphics[width=0.49\linewidth]{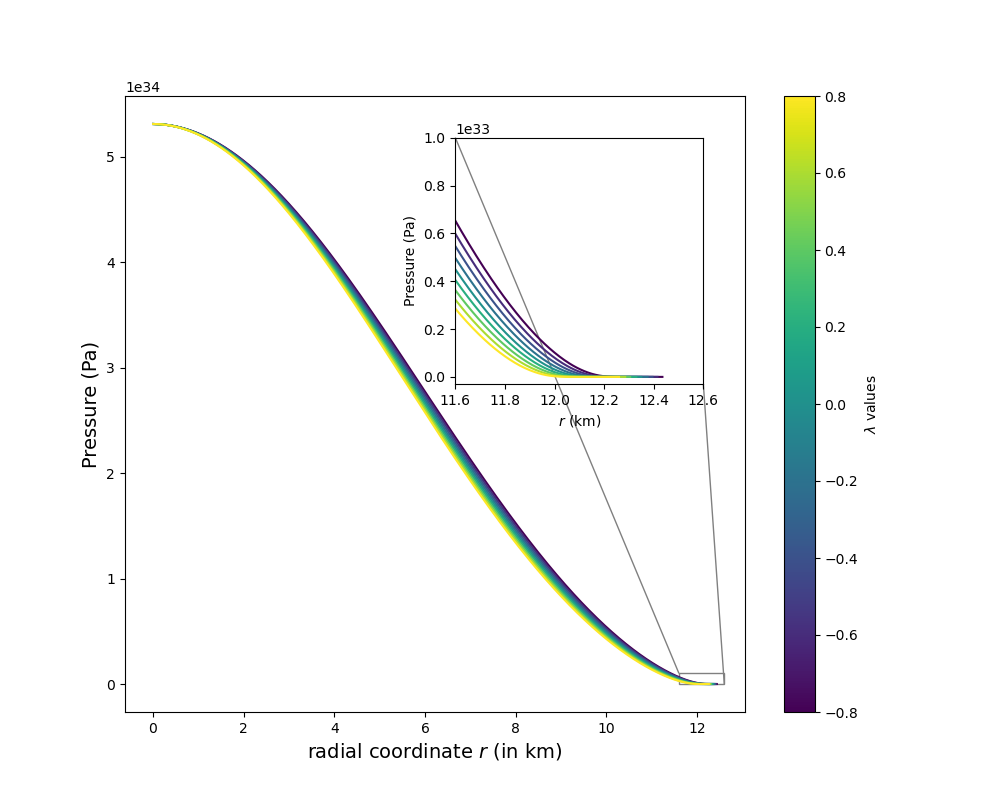}\label{fig:DD2_PVRsingle}}

\subfigure[\ m$(M_\odot)$  vs $r$ (in km) of DDH$_\delta$ EOS]
{\includegraphics[width=0.49\linewidth]{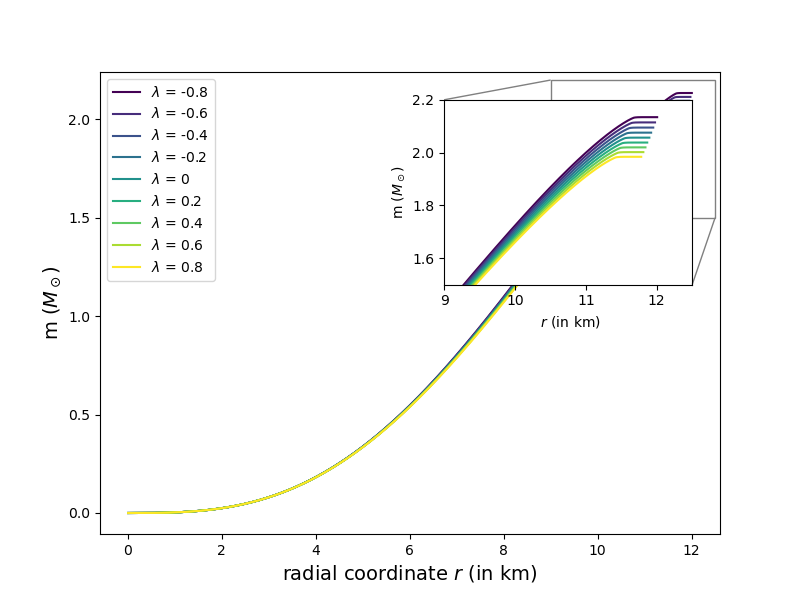}\label{fig:DDH_MVRsingle}} 
\subfigure[\ Pressure(Pa)  vs $r$ (in km) of DDH$_\delta$ EOS]{\includegraphics[width=0.49\linewidth]{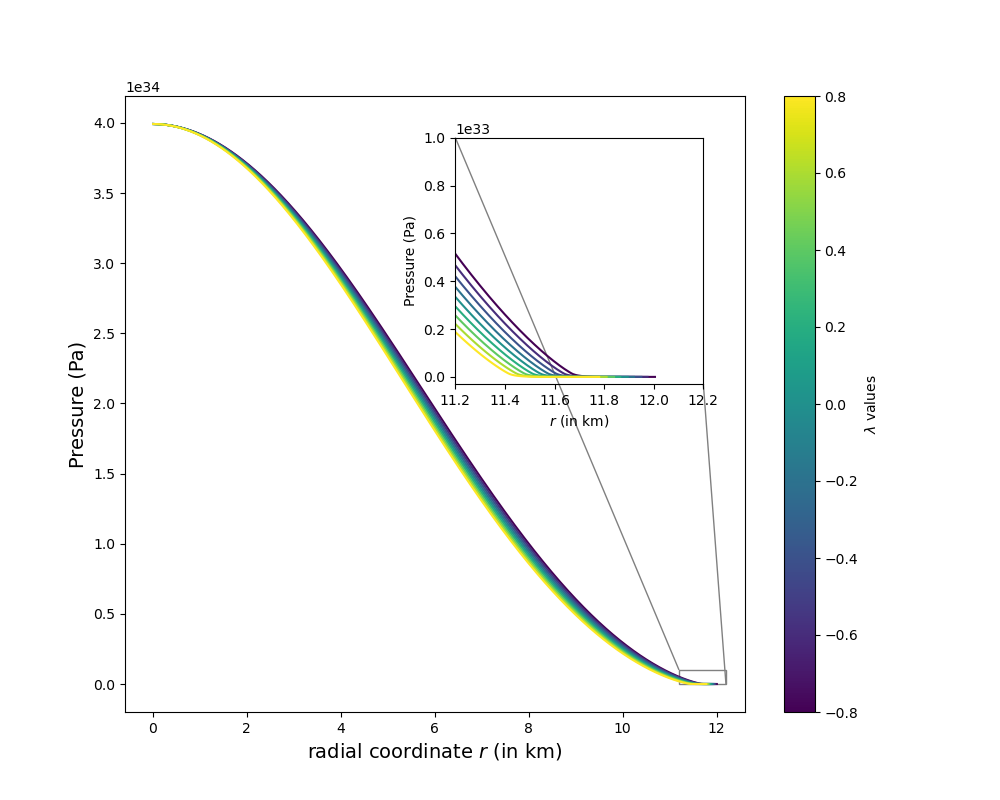}\label{fig:DDH_PVRsingle}}
\subfigure[\ m$(M_\odot)$  vs $r$ (in km) of TW EOS]{\includegraphics[width=0.49\linewidth]{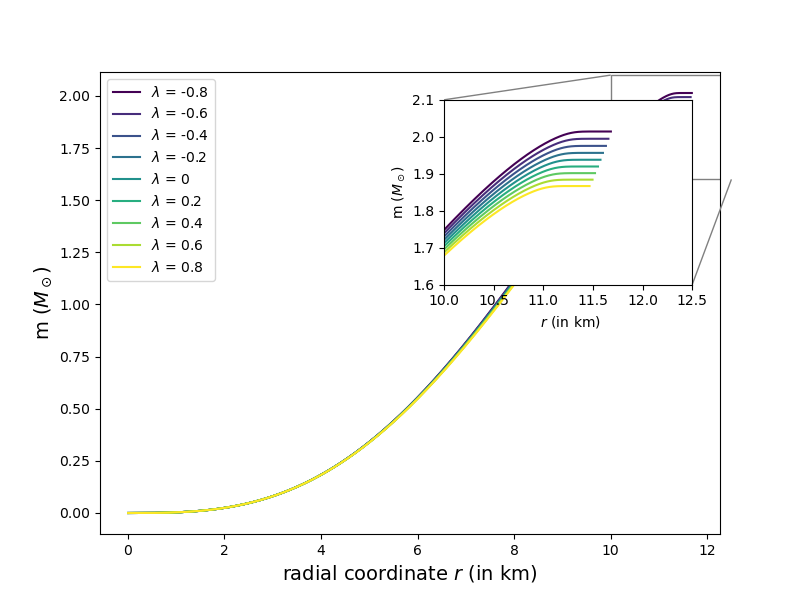}\label{fig:TW_MVRsingle}} 
\subfigure[\ Pressure(Pa)  vs $r$ (in km) of TW EOS]{\includegraphics[width=0.49\linewidth]{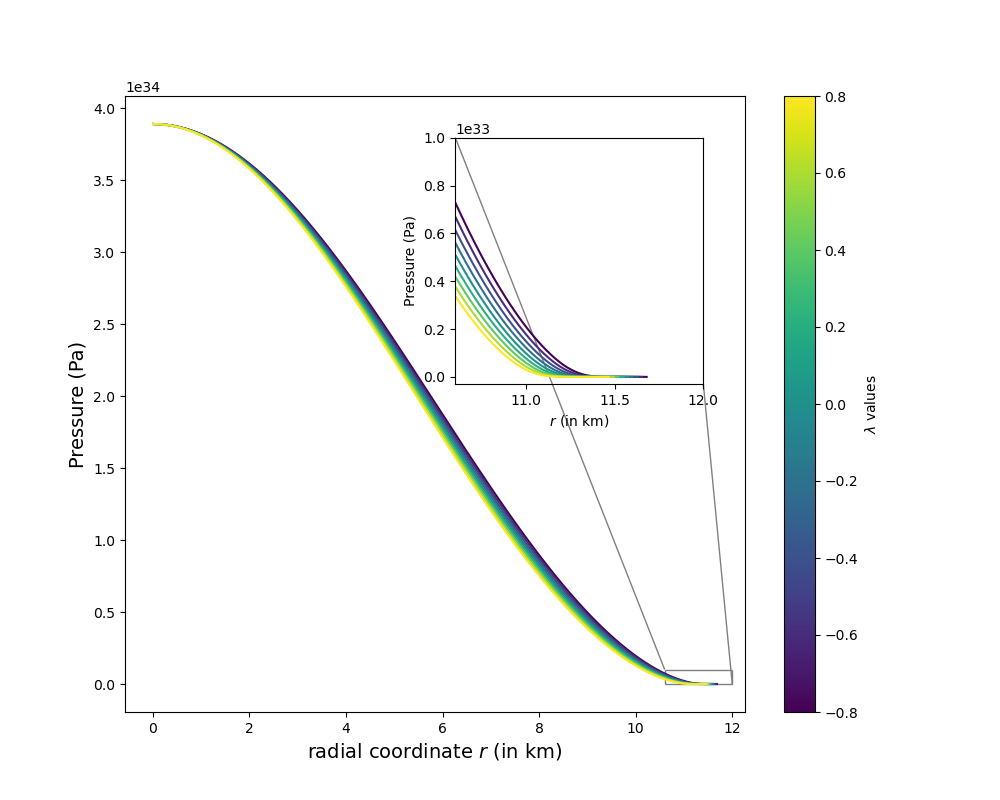}\label{fig:TW_PVRsingle}}
\caption{\justifying Mass vs radial coordinate ($r$)  and Pressure vs radial coordinate  ($r$) profile for a single Neutron star from the core to the surface of the star \textbf{DD2, DDH$_\delta$, TW EOS}}
\label{1ST3rmf_SINGLE}
\end{figure}

\end{widetext}

\begin{widetext}

\begin{figure}[H]
\centering
\subfigure[\ m$(M_\odot)$  vs $r$ (in km) of NL3 EOS]
{\includegraphics[width=0.49\linewidth]{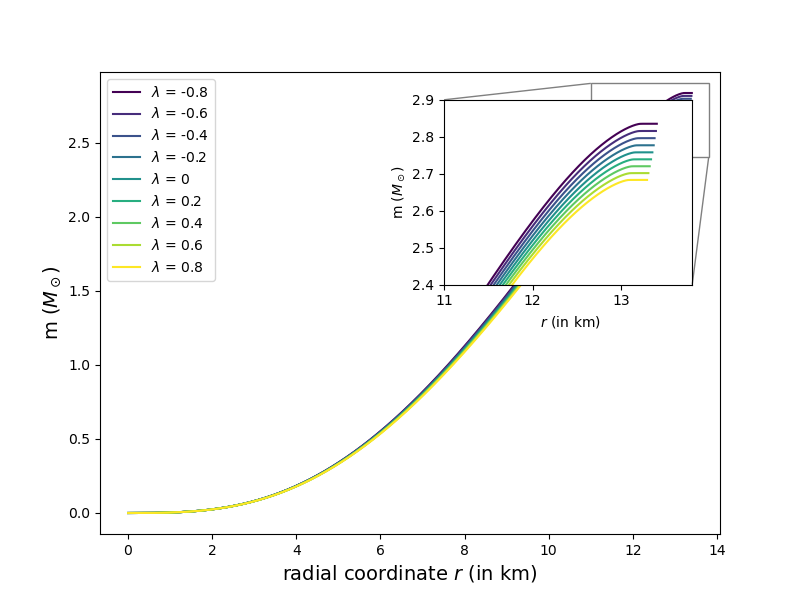}\label{fig:NL3_MVRsingle}} 
\subfigure[\ Pressure(Pa)  vs $r$ (in km) of NL3 EOS]{\includegraphics[width=0.49\linewidth]{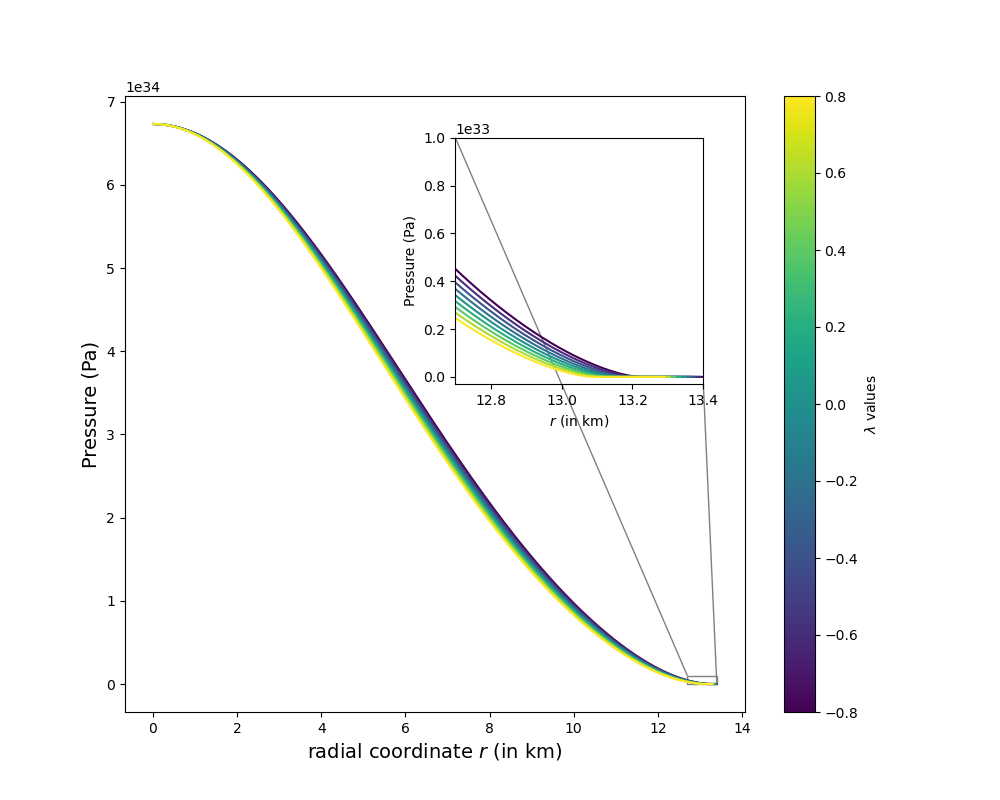}\label{fig:NL3_PVRsingle}}
\subfigure[\ m$(M_\odot)$  vs $r$ (in km) of GM1 EOS]
{\includegraphics[width=0.49\linewidth]{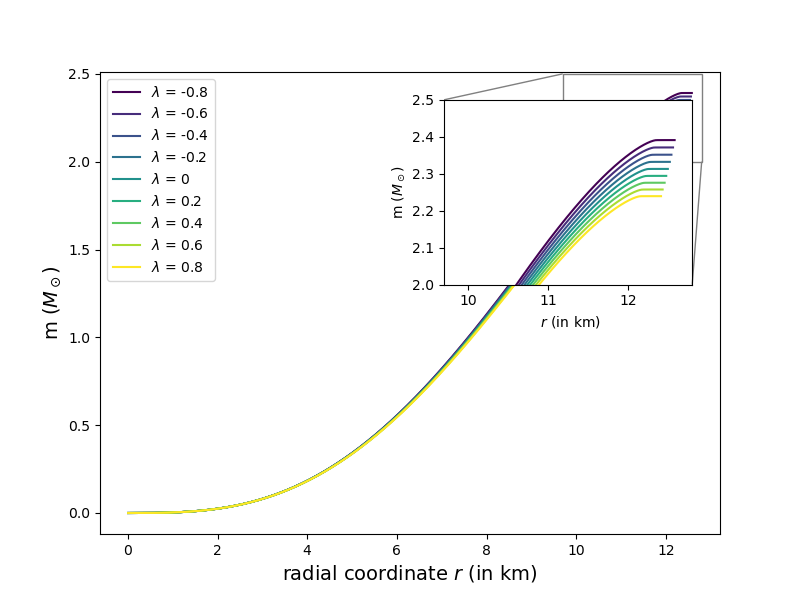}\label{fig:GM1_MVRsingle}} 
\subfigure[\ Pressure(Pa)  vs $r$ (in km) of GM1 EOS]{\includegraphics[width=0.49\linewidth]{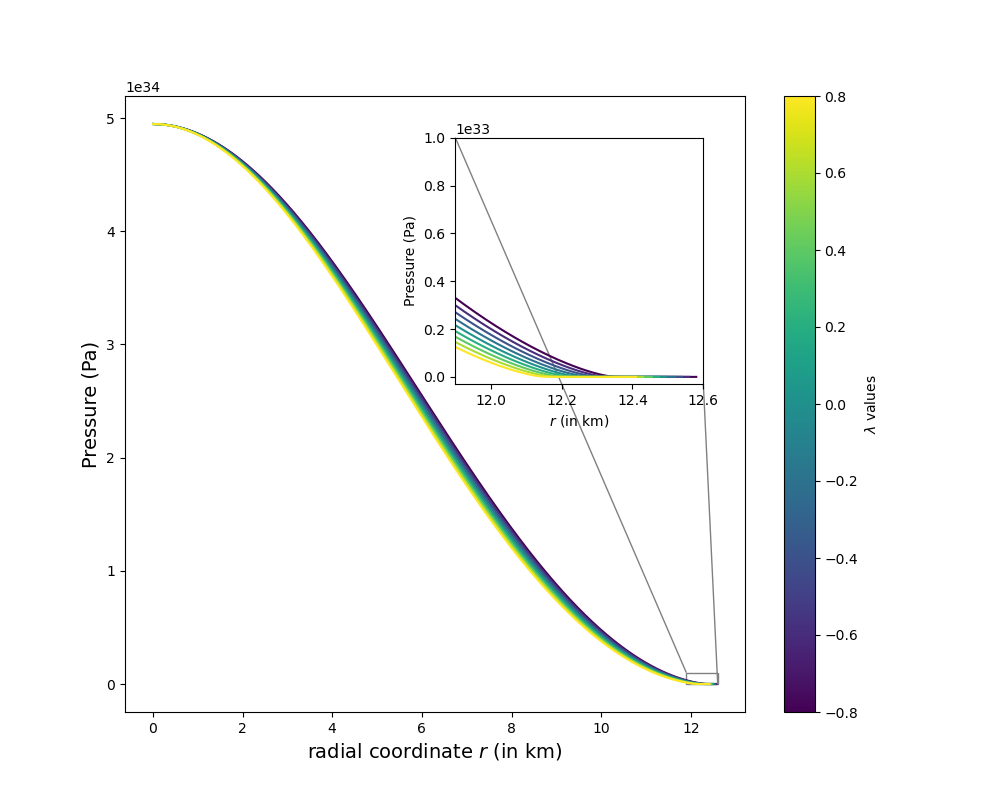}\label{fig:GM1_PVRsingle}}
\subfigure[\ m$(M_\odot)$  vs $r$ (in km) of TM1 EOS]{\includegraphics[width=0.49\linewidth]{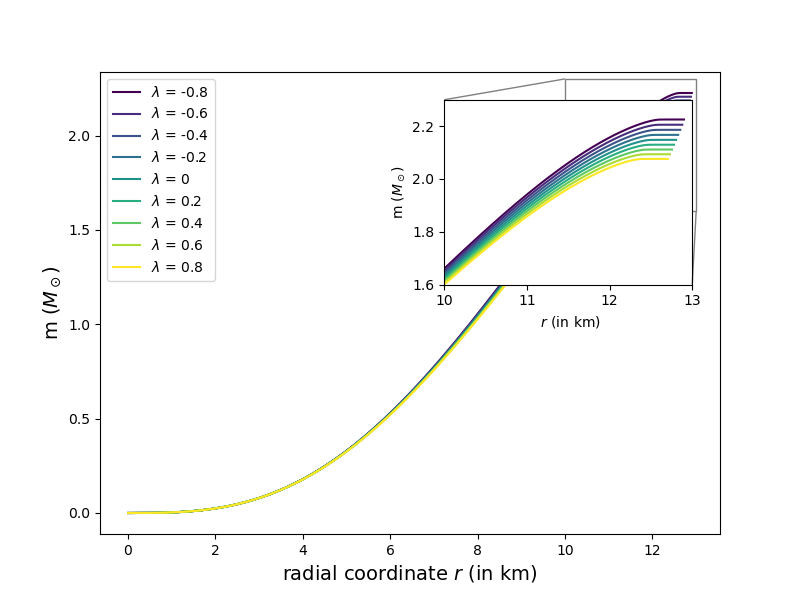}\label{fig:TM1_MVRsingle}} 
\subfigure[\ Pressure(Pa)  vs $r$ (in km) of TM1 EOS]{\includegraphics[width=0.49\linewidth]{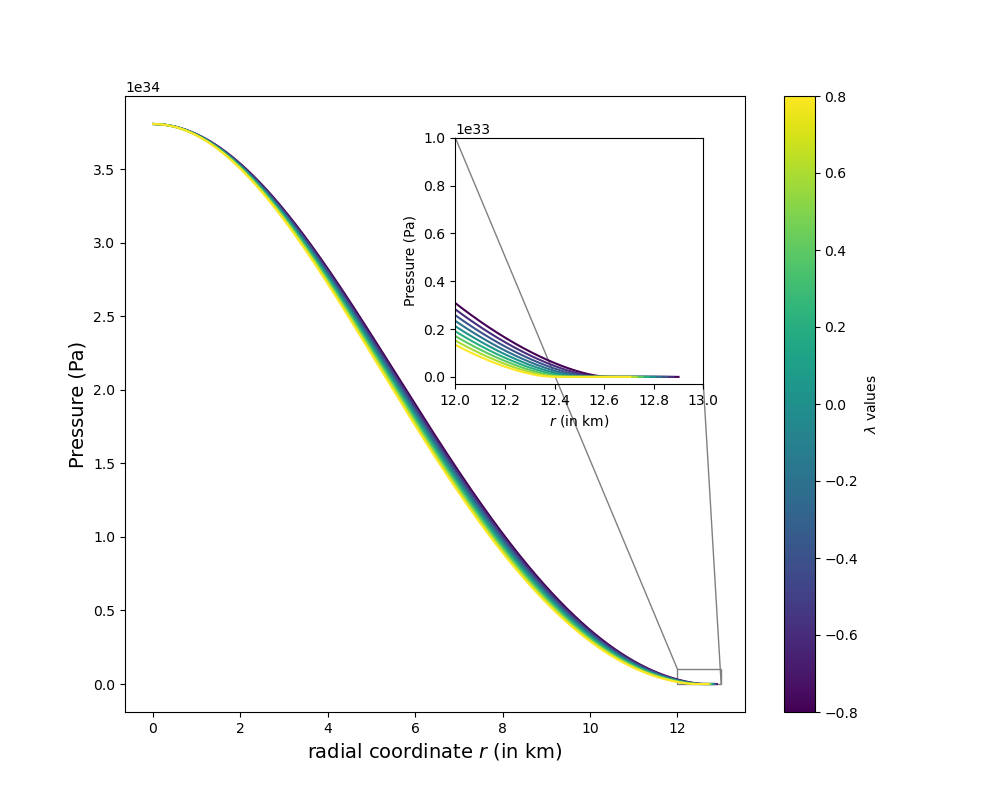}\label{fig:TM1_PVRsingle}}
\caption{\justifying Mass vs radial coordinate ($r$)  and Pressure vs radial coordinate  ($r$) profile for a single Neutron star from the core to the surface of the star \textbf{NL3, GM1, TM1 EOS}}
\label{LAST3rmf_SINGLE}
\end{figure}

\end{widetext}

Although there are several studies in minimally coupled models, we started our analysis taking this conserved $f(R, T) = R +  \lambda T $ model,  \cite{Lobato:2020fxt, Carvalho:2017pgk,dosSantos:2018nmu, Pretel:2020oae, Pretel:2022qng}. We solve the TOV equations using RMF-based realistic EOS (as mentioned above) for neutron star matter corresponding to different $\lambda$ values. In this work, we have chosen $\lambda \sim \mathcal{O}(10^{-1}$ or $10^{-2}$) which satisfy the causality condition inside the neutron star. As a star satisfies the hydrostatic equilibrium condition, at the surface the neutron degeneracy pressure is not that much sufficient to balance the gravity, signifying pressure is zero at the surface of a star where the mass $m$ (in $M\odot$ unit) will be the maximum. We have shown Pressure $P$ (in Pa) against the radial coordinate($r$)in the interior of the single NS  corresponding to the values of  $\lambda = -0.8, -0.6,-0.4,-0.2,0.0, 0.2,0.4.0.6$ and $0.8$ \footnote{ Note that $\lambda = 0$ corresponds to normal GR.} 
for a particular value of central energy density $\rho_c = 1.5 \times 10^ {15} g /cm^3$ for all the RMF EOS. The value of radial coordinate $r$ where it is becoming the maximum known as Radius $R$ of the single star (also where Pressure; which is order of $10^{34}$ Pa, vanishes) is zoomed out inside the plots as shown in fig. (\ref{fig:DD2_PVRsingle}, \ref{fig:DDH_PVRsingle}, \ref{fig:TW_PVRsingle},\ref{fig:NL3_PVRsingle},\ref{fig:GM1_PVRsingle},  \ref{fig:TM1_PVRsingle}) for all EOS. Starting from the 3 density-dependent models, for the DD2 model, the maximum mass is   $2.36 M_\odot$ with the radius of the single star found to be $12.35$ km.

When we try to increase the values of $\lambda$ positively (and negatively) up to $+0.8$ (and $-0.8$) the maximum mass decreases (and increases) up to $2.29$ (and $2.44) M_\odot$ with radius $12.26$ (and $12.43$) km where pressure vanishes as shown in figure (\ref{fig:DD2_MVRsingle}).

Similarly for DDH$_\delta$ and TW EOS  the maximum masses are  $2.05$ (and $1.93) M_\odot$ and the radius is $11.89$ (and $11.57$) km. While increasing (and decreasing) the values of $\lambda$ up to $+0.8$ (and $-0.8$) we get values up to maximum mass of $1.98$ (and $2.13) M_\odot$ \& radius of   $11.78$ (and $12.00$) km for  DDH$_\delta$ EOS and maximum mass of $1.86$ (and $2.01) M_\odot$ \& radius of   $11.46$ (and $11.68$) km for TW EOS respectively as shown in the figures (\ref{fig:DDH_MVRsingle} and \ref{fig:TW_MVRsingle})

 The non-linear interacting models with self-coupling like NL3 show maximum mass to be $2.75 M\odot$ and the radius to be $13.35$ km. The other 2 models i.e. GM1 and TM1 also show the maximum mass $2.31$ (and $2.14) M_\odot$ \& radius $12.49$ (and $12.80$) km at the surface of the single neutron star respectively.

\begin{widetext}

\begin{figure}[H]
\centering
\subfigure[\ Mass$(M_\odot)$  vs Radius $R$ (in km) ]{\includegraphics[width=0.49\linewidth]{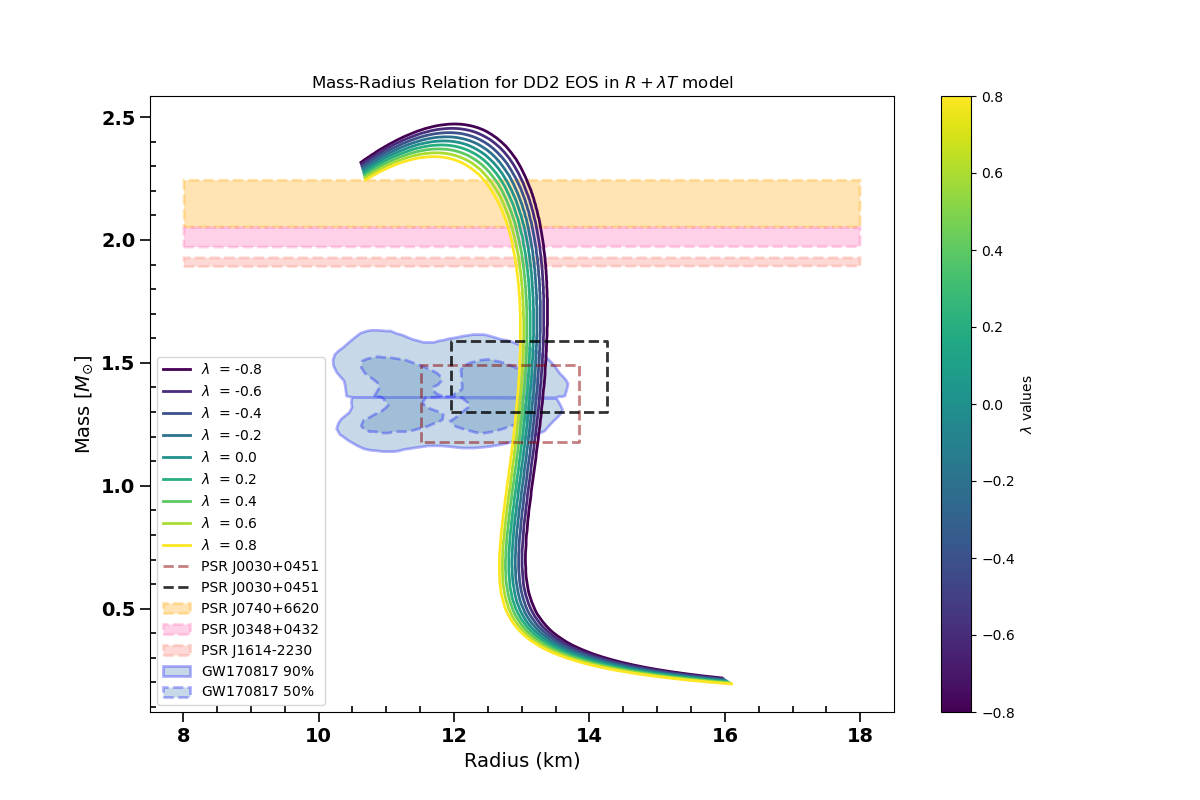}\label{fig:DD2_MVRnumberofstars}} 
\subfigure[\ Mass$(M_\odot)$  vs Radius $R$ (in km) ]{\includegraphics[width=0.49\linewidth]{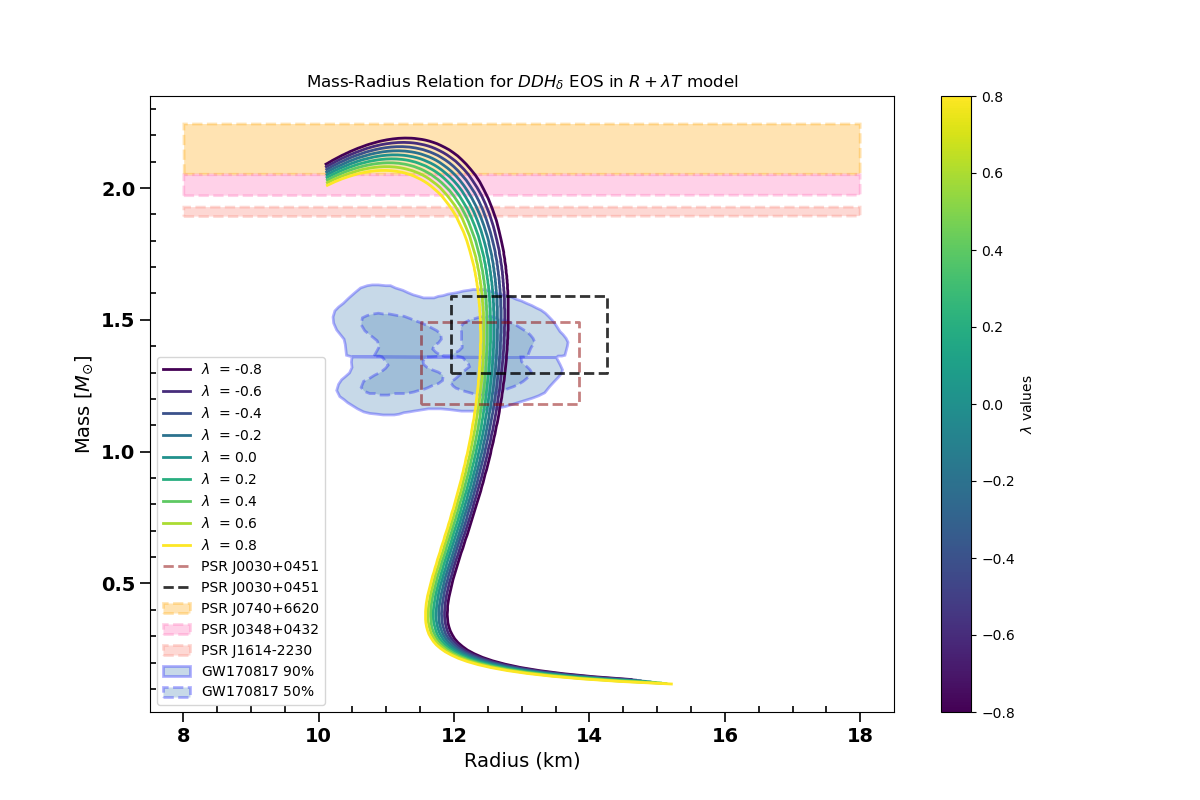}\label{fig:DDH_MVRnumberofstars}}
\subfigure[\ Mass$(M_\odot)$  vs Radius $R$ (in km) ]{\includegraphics[width=0.49\linewidth]{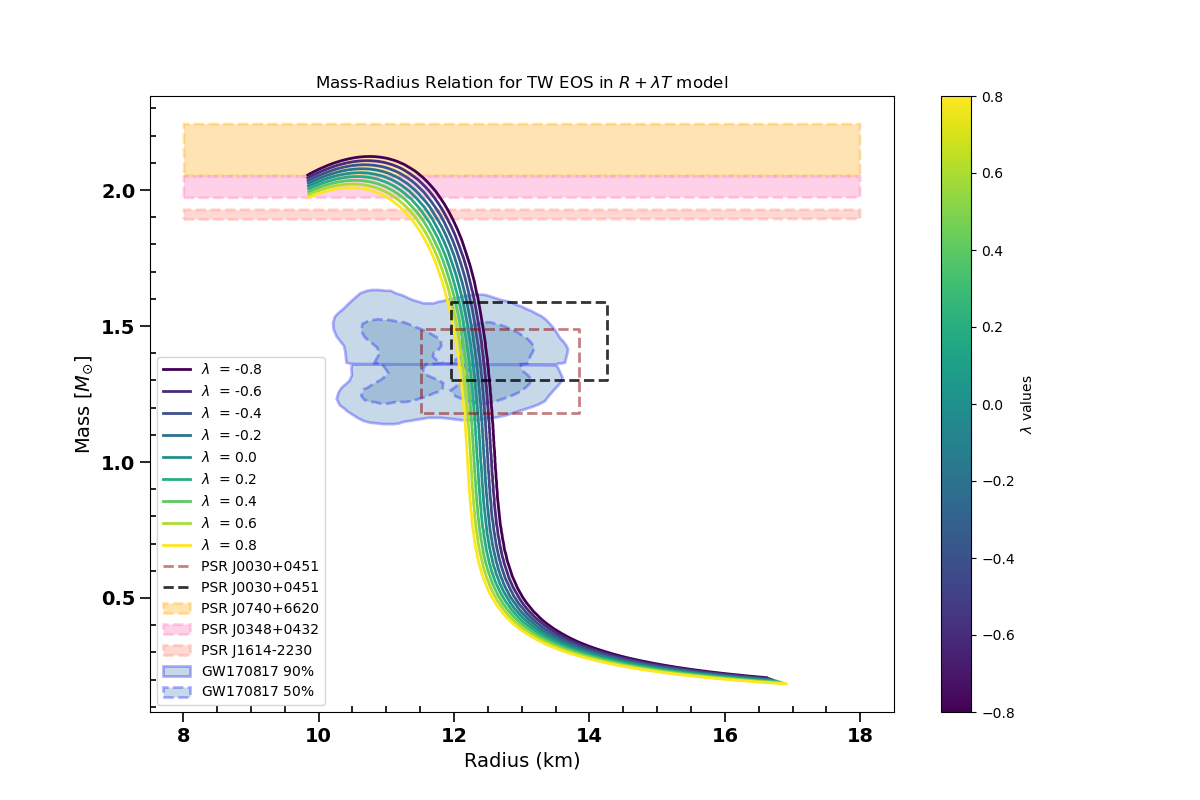}\label{fig:TW_MVRnumberofstars}} 
\subfigure[\ Mass$(M_\odot)$  vs Radius $R$ (in km) ]{\includegraphics[width=0.49\linewidth]{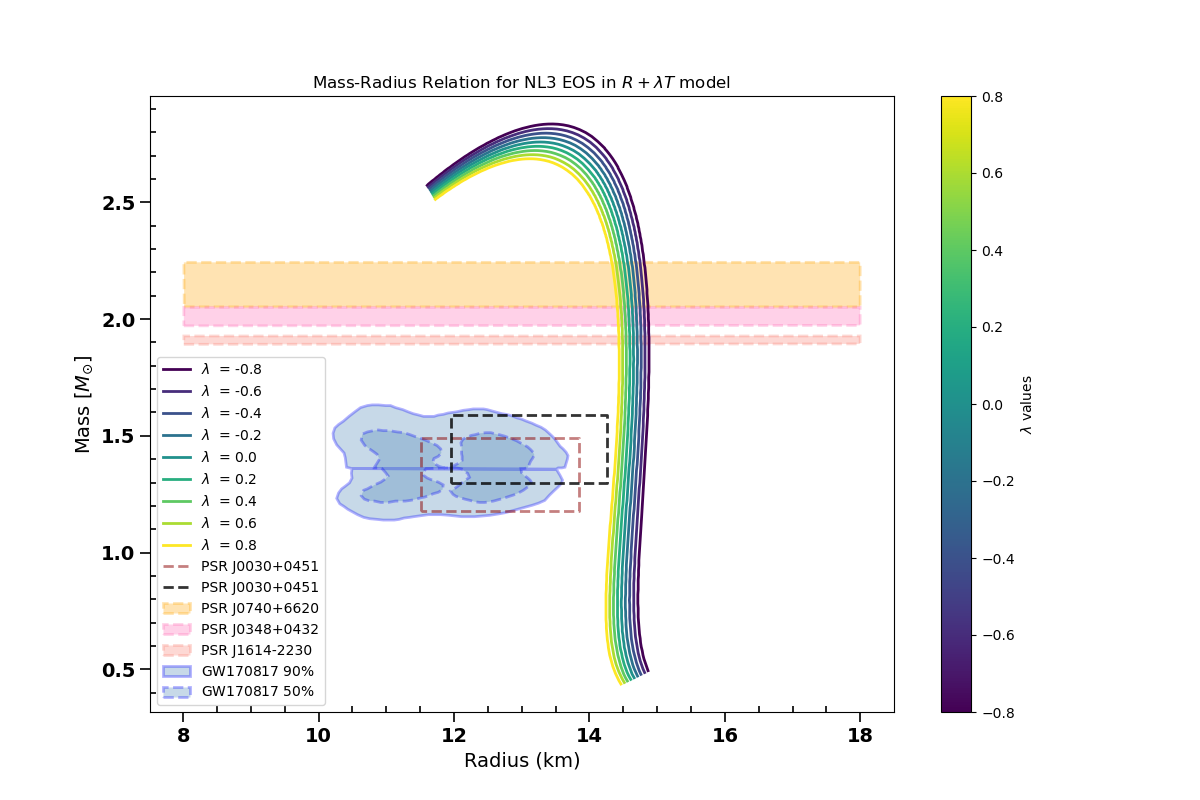}\label{fig:NL3_MVRnumberofstars}} 
\subfigure[\ Mass$(M_\odot)$  vs Radius $R$ (in km) ]{\includegraphics[width=0.49\linewidth]{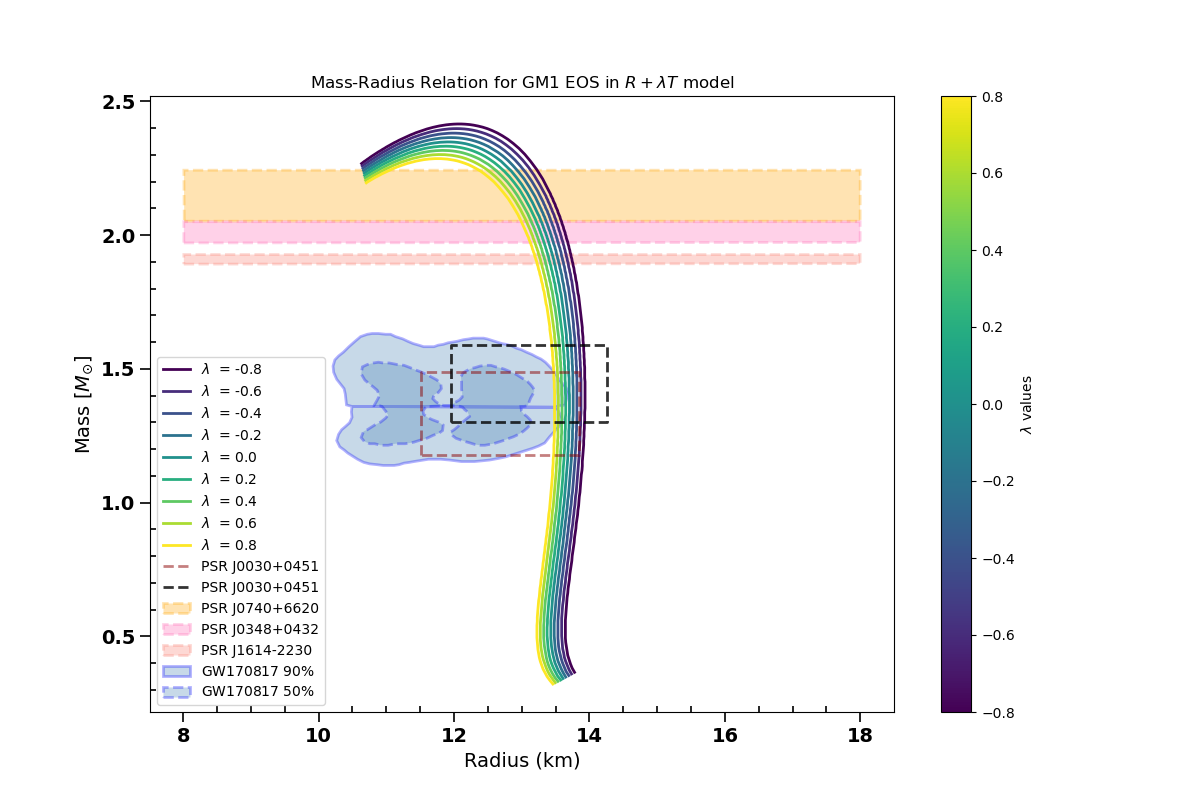}\label{fig:GM1_MVRnumberofstars}} 
\subfigure[\ Mass$(M_\odot)$  vs Radius $R$ (in km) ]{\includegraphics[width=0.49\linewidth]{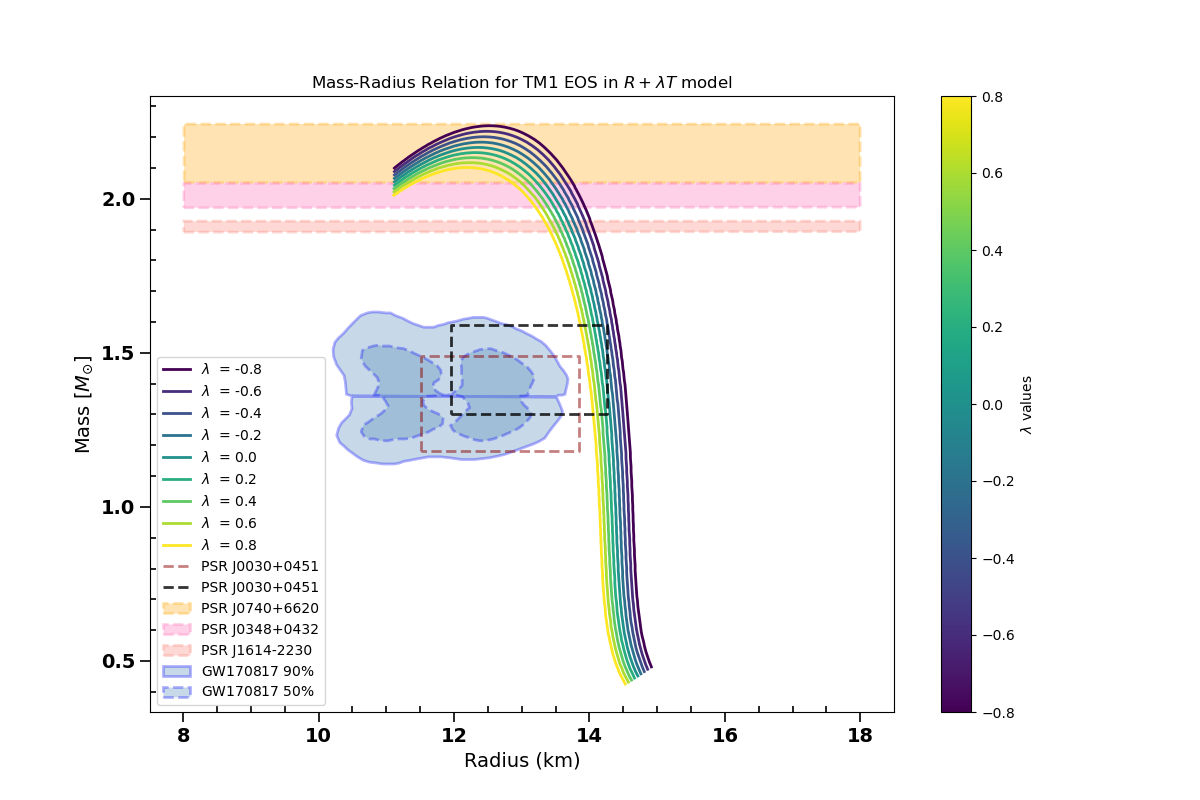}\label{fig:TM1_MVRnumberofstars}} 

\caption{\justifying Mass($M_\odot$) vs Radius ($R$) for a number of Neutron Stars in the given range of $\rho_c$ for \textbf{DD2, DDH$_\delta$, TW, NL3, GM1, TM1 EOS}} 
\label{LAMBDAT_ALL}
\end{figure}

\end{widetext}

 Also the positive ( and negative) values of $\lambda$ of the order of $\smash{\scriptstyle\pm} 0.2$ up to $\smash{\scriptstyle\pm} 0.8$ impact their maximum mass and radius like  $2.68$ (and $2.83) M_\odot$, $2.23$ (and $2.39) M_\odot$, $2.07$ (and $2.22) M_\odot$ \&  $13.29$ (and $13.4$) km,  $12.41$ (and $12.58$) km,  $12.70$ (and $12.90$) km for NL3, GM1, and TM1 EOS as shown in figure (\ref{fig:NL3_MVRsingle}, \ref{fig:GM1_MVRsingle} and \ref{fig:TM1_MVRsingle}) respectively, where the pressure vanishes at the surface.

Now, we will see how for a range of central energy density $\rho_c  = 2.5 \times 10^{14} g /cm^3$ to $4 \times 10^{15} g /cm^3 $ when \eqref{dmdrMod} and \eqref{dpdrMod} are solved, we got the Mass ($M_\odot$) vs Radius curves of several neutron stars in that range for different values of $\lambda$.

Depending on the \textit{soft} or \textit{stiff} EOS, based on the compressibility of nuclear matter and their response at high densities i.e. (how fast the pressure changes when the energy density changes), the maximum mass $M_{max}$, maximum radius $R_{max}$ and radius at 1.4 $M\odot$ i.e. $R_{1.4}$ for all the 6 RMF models are described in table (\ref{tab:eos_data}).

As shown in figure (\ref{fig:NL3_MVRnumberofstars}, \ref{fig:GM1_MVRnumberofstars}, \ref{fig:TM1_MVRnumberofstars}), and table (\ref{tab:eos_data})  the nonlinear interacting 3 EOS models NL3, GM1 and TM1 show the maximum mass of $2.75, 2.34$ and $2.16 M_\odot$ with $R_{1.4} = 14.59, 13.70$ and $14.25$ km for $\lambda = 0$ in GR. When subjected to different modified $\lambda$ values (both positive and negative), it gives a result that is \textit{\textbf{not satisfying}} the PSR J0030+0451 maximum mass - radius region given by NICER x-ray data and GW170817 $90 \%$ and $50 \% $ CI region.

Among the density-dependent models like DDH$_\delta$ and TW EOS pass through all of the given observational constraints like GW170817 radius range of $10.62$ km $< R_{1.4} < 12.83$ km  \cite{Nathanail:2021tay, Shibata:2019ctb, abbott2018gw170817, GWOSC_softx}, the millisecond pulsar PSR J0030+0451 (black \& dark red rectangular region) NICER x-ray data \cite{Miller:2019cac, Riley:2019yda} and all the 3 given pulsars PSRJ1614–22 \cite{Takisa:2014sva}, PSR J0348+0432 \cite{Zhao:2016rfv} and  PSR J0740+6620 \cite{Miller:2021qha, Riley:2021pdl} maximum mass limit for   $\lambda = 0.4, 0.6$ and $-0.4, -0.2$ respectively. But DD2 EOS just cross the NICER x-ray data and $90 \%$ CI region of GW170817 data, not both regions giving the maximum mass of $2.33  M_\odot$ for $\lambda =0.8$ as shown in figures (\ref{fig:DDH_MVRnumberofstars}, \ref{fig:TW_MVRnumberofstars} and \ref{fig:DD2_MVRnumberofstars}), respectively and the clear information is in table (\ref{tab:eos_data}).

\begin{widetext}
    
\begin{figure}[H]
\centering
\subfigure[ $c_s^2 / c^2$ vs Number density of DD$2$ EOS]{\includegraphics[width=0.49\linewidth]{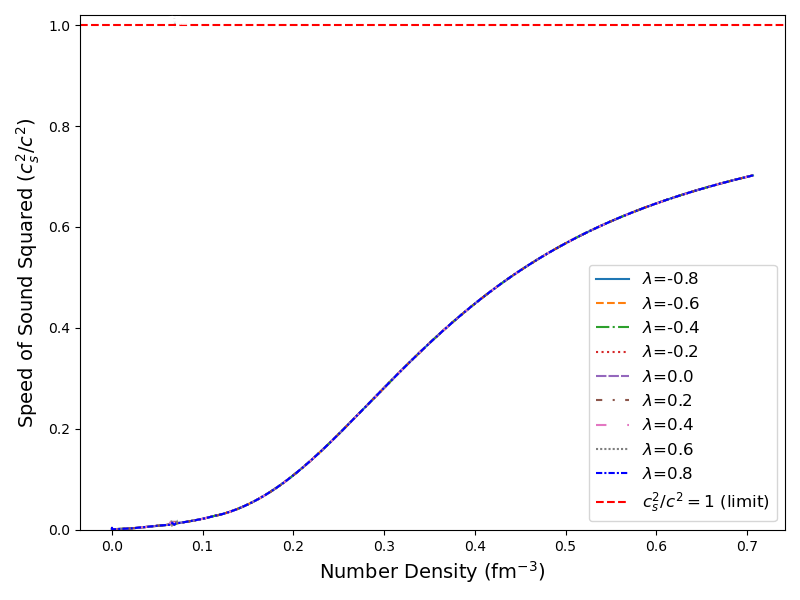}\label{fig:DD2_soundspeedlambda}} 
\subfigure[$c_s^2 / c^2$ vs Number density of DDH$_\delta$ EOS ]{\includegraphics[width=0.49\linewidth]{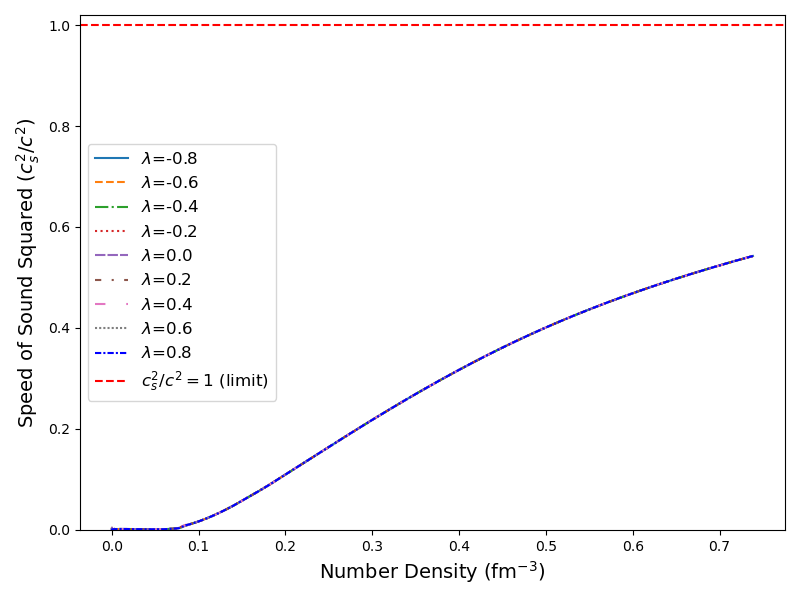}\label{fig:DDH_soundspeedlambda}}
\subfigure[$c_s^2 / c^2$ vs Number density of TW EOS ]{\includegraphics[width=0.49\linewidth]{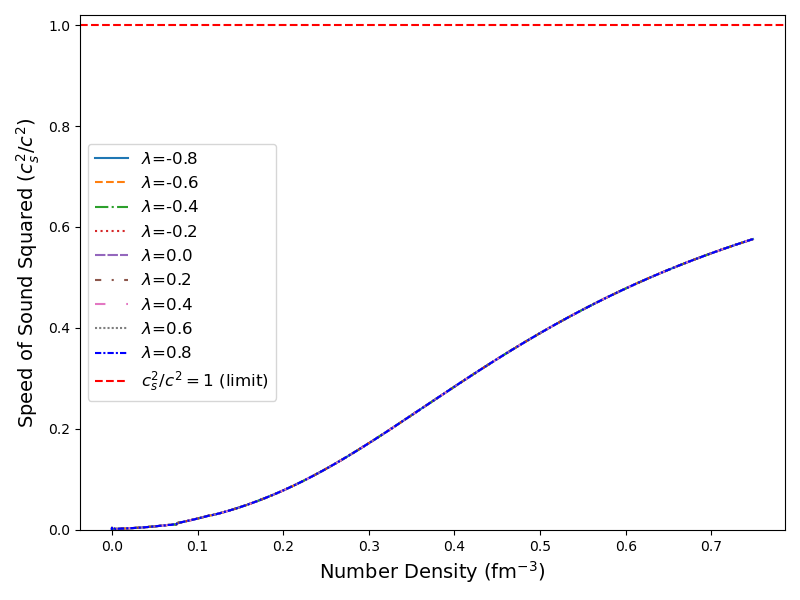}\label{fig:TW_soundspeedlambda}} 
\subfigure[$c_s^2 / c^2$ vs Number density of NL$3$ EOS ]{\includegraphics[width=0.49\linewidth]{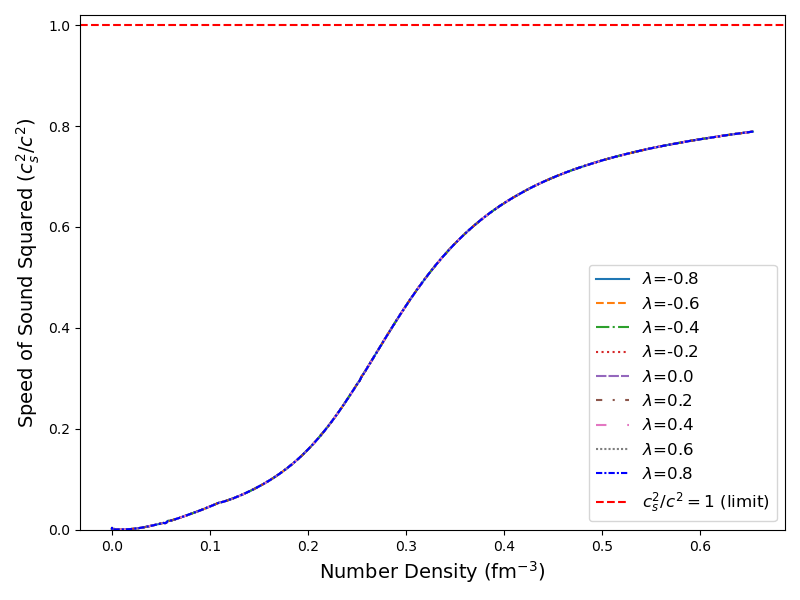}\label{fig:NL3_soundspeedlambda}} 
\subfigure[$c_s^2 / c^2$ vs Number density of GM$1$ EOS ]{\includegraphics[width=0.49\linewidth]{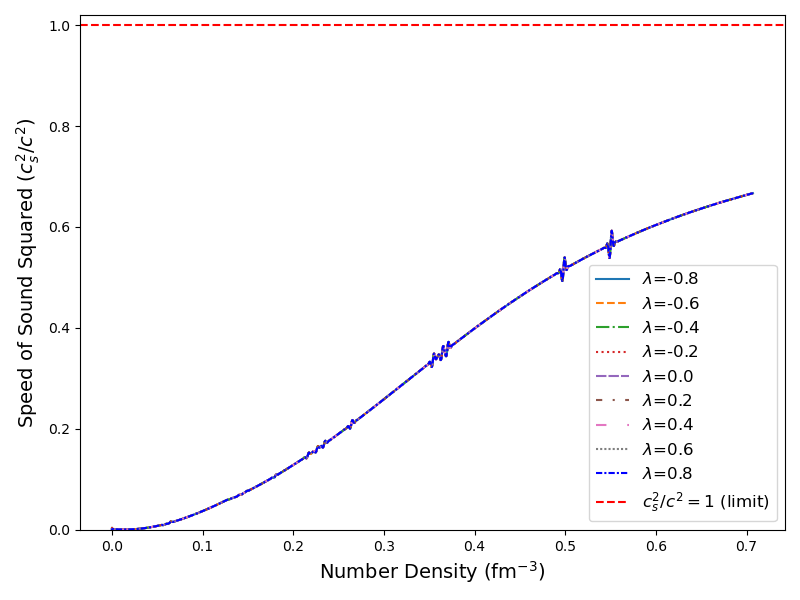}\label{fig:GM1_soundspeedlambda}} 
\subfigure[$c_s^2 / c^2$ vs Number density of TM$1$ EOS ]{\includegraphics[width=0.49\linewidth]{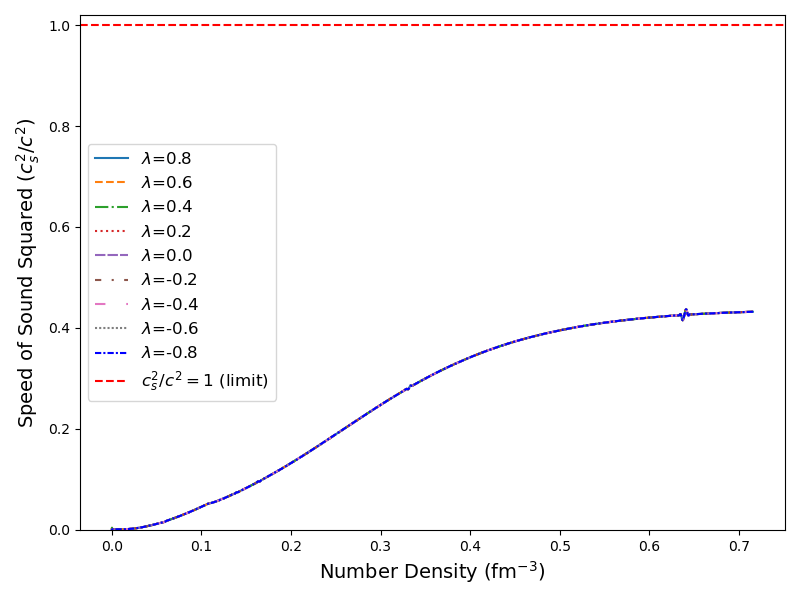}\label{fig:TM1_soundspeedlambda}} 

\caption{\justifying Speed of Sound Squared ($c_s^2 / c^2$) vs Number Density ($\mathrm{fm}^{-3}$) for all the given RMF EOS viz. \textbf{DD2, DDH$_\delta$, TW, NL3, GM1, TM1 }} 
\label{speedofsoundmodified}
\end{figure}
\end{widetext}

Taking these modified TOV equations (\ref{dmdrMod}) \&(\ref{dpdrMod}), the outcome Pressure which also depends upon the extra term $h(T)$ of our $f(R, T)$ model and also significantly change the behaviour of mass-radius relation along with the speed of sound got changed (stiffness of the EOS). As the speed of sound squared $c_{s}^{2}$ is dependent on the gradient of Pressure, it will be changed for different values of $\lambda$. As shown in the figure (\ref{speedofsoundmodified}) and table (\ref{tab:maximum-cs2}), the change in speed of sound squared $c_{s}^{2} = dP / d \rho$ for different value of $\lambda$ is very very less, which got changed after $6th$ decimal point as compared to the GR result (for $\lambda = 0$). The modification in the quantity $c_{s}^{2}$ is so small that, all the curves look alike in figure (\ref{speedofsoundmodified}) for all taken 6 RMF models. Taking negative(positive) values of $\lambda$, the speed of sound squared increased(decreased) for all models and all of them are below the causality limit ($c_s^2/c^2 =1$), satisfying this condition necessary for stability.

The behaviour of the total mass of all these RMF EOS against the central energy density range $\rho_c = 2.5 $ to $ 4 \times 10^{15} g /cm^3$ is shown in figure (\ref{LAMBDAT_ALL-RHO}) for all the 6 models in the appendix~\ref{mass-rhoccurve}. It is remarkable that for lower values of $\lambda$, the maximum mass point for all the EOS is reached for the lower value of $\rho_c$, after that point the stellar mass decreases with the increment of central energy density value $\rho_c$.

\begin{widetext}
    \begin{table}[!htb]
    \centering
    \scriptsize 
    \setlength{\tabcolsep}{3pt} 
    \captionsetup{width=\textwidth}
    \resizebox{\textwidth}{!}{%
    \begin{tabular}{|c|c|c|c|c|c|c|c|c|c|c|c|c|c|c|c|c|c|c|}
        \hline
        \multirow{2}{*}{\textbf{$\lambda$}} & \multicolumn{3}{c|}{\textbf{DD$2$}} & \multicolumn{3}{c|}{\textbf{DDH$_\delta$}} & \multicolumn{3}{c|}{\textbf{TW}} & 
        \multicolumn{3}{c|}{\textbf{NL$3$}} & \multicolumn{3}{c|}{\textbf{GM$1$}} & \multicolumn{3}{c|}{\textbf{TM$1$}} \\
        \cline{2-19}
        & \textbf{$M_{max}$} & \textbf{$R_{max}$} & \textbf{$R_{1.4}$} & \textbf{$M_{max}$} & \textbf{$R_{max}$} & \textbf{$R_{1.4}$} & \textbf{$M_{max}$} & \textbf{$R_{max}$} & \textbf{$R_{1.4}$} & \textbf{$M_{max}$} & \textbf{$R_{max}$} & \textbf{$R_{1.4}$} & \textbf{$M_{max}$} & \textbf{$R_{max}$} & \textbf{$R_{1.4}$} & \textbf{$M_{max}$} & \textbf{$R_{max}$} & \textbf{$R_{1.4}$} \\
        \hline
        0.0 (GR) & 2.4029 & 11.855 & 13.1317 & 2.1259 & 11.13 & \textbf{12.585} & 2.0638 & 10.6 & \textbf{12.2618} & 2.7583 & 13.275 & 14.5977 & 2.3485 & 11.9 & 13.705 & 2.1669 & 12.34 & 14.2589  \\
        
        -0.8 & 2.4717 & 12.01 & 13.3265 & 2.1896 & 11.275 & \textbf{12.7856} & 2.1236 & 10.75 & \textbf{12.4804} &  2.8354 & 13.46 & 14.8217 & 2.4156 & 12.05 & 13.935 & 2.2373 & 12.515 & 14.5234 \\
        
        -0.6 & 2.4540 & 11.945 & 13.2741 & 2.1732 & 11.26 & \textbf{12.735} & 2.1082 & 10.7 & \textbf{12.4247} & 2.8156 & 13.375 & 14.7657 & 2.3984 & 12.04 & 13.875 &  2.2192 & 12.495 & 14.4534 \\
        
        -0.4 & 2.4366 & 11.93 & 13.2299 & 2.1571 & 11.2 & \textbf{12.685} & \textbf{2.0931} & 10.69 & \textbf{12.3688} & 2.7961 & 13.365 & 14.7097 & 2.3814 & 11.97 & 13.815 & 2.2014 & 12.425 & 14.3892  \\
        
        -0.2 & 2.4196 & 11.87 & 13.1808 & 2.1413 & 11.185 & \textbf{12.635} & \textbf{2.0783} & 10.645 & \textbf{12.315} & 2.7769 & 13.285 & 14.6524 & 2.3648 & 11.96 & 13.76 &  2.1840 & 12.405 & 14.3216 \\

        0.2 & 2.3865 & 11.8 & 13.0877 & 2.1107 & 11.075 & \textbf{12.535} & 2.0496 & 10.59 & \textbf{12.2106} & 2.7398 & 13.26 & 14.5416 & 2.3326 & 11.885 & 13.6490 &  2.1501 & 12.32 & 14.1924 \\
        
        0.4 & 2.3704 & 11.785 & 13.0387 & \textbf{2.0958} & 11.06 & \textbf{12.4861} & 2.0357 & 10.545 & \textbf{12.1592} & 2.7218 & 13.19 & 14.4856 & 2.3169 & 11.83 & 13.5919 &  2.1337 & 12.255 & 14.1318  \\
       
        0.6 & 2.3546 & 11.73 & 12.9947 & \textbf{2.0812} & 11.005 & \textbf{12.44} & 2.0220 & 10.5 & \textbf{12.1078} & 2.7040 & 13.175 & 14.4345 & 2.3014 & 11.815 & 13.5399 & 2.1176 & 12.235 & 14.0705 \\
        
        0.8 & 2.3390 & 11.72 & 12.9508 & 2.0669 & 10.955 & \textbf{12.39} & 2.0085 & 10.46 & \textbf{12.0562} & 2.6865 & 13.11 & 14.3834 & 2.2864 & 11.76 & 13.485 &  \textbf{2.1017} & 12.175 & 14.0076 \\
        \hline
    \end{tabular}}
    \caption{Maximum mass $M_{\max}$ (in $M_\odot$), maximum radius $R_{\max}$ (in km), and radius at 1.4 $M_{\odot}$ i.e $R_{1.4}$ (in km) for different EOS files (DD$2$, DDH$_\delta$, TW, NL$3$, GM$1$, TM$1$) as a function of $\lambda$.}
    \label{tab:eos_data}
    \end{table}

\end{widetext}

\subsection{Maximum mass of neutron stars in $f(R,T)$ gravity}
\label{subsec:maxmass}

{\color{black}In the preceding subsections, we have presented the numerical solutions of the modified Tolman--Oppenheimer--Volkoff equations for a wide class of relativistic mean-field equations of state within conservative $f(R,T)$ gravity. While mass--radius relations and internal stellar profiles provide important insights into the role of matter--geometry coupling, the maximum gravitational mass of neutron stars constitutes one of the most stringent and observationally robust constraints on both dense-matter microphysics and the underlying theory of gravity.

Motivated by recent studies emphasizing the diagnostic power of maximum mass bounds in extended gravity theories \cite{Astashenok:2021btj, Astashenok:2021peo}, we now focus on a detailed discussion of the maximum mass of neutron stars in conservative $f(R, T)$ gravity. In particular, we analyze how the coupling parameter $\lambda$ influences the maximum stable configuration and how these effects interplay with the stiffness of realistic RMF equations of state under current multi-messenger observational constraints.

The maximum gravitational mass of neutron stars represents one of the most robust and observationally constrained quantities in compact-star physics. Precise mass measurements of heavy pulsars, such as PSR~J1614--2230 and PSR~J0740+6620, as well as constraints inferred from the binary neutron star merger GW170817, provide stringent lower bounds on the maximum mass that any viable equation of state (EOS) and gravitational theory must satisfy. At the same time, the maximum mass is highly sensitive to modifications of gravity in the strong-field regime, making it a powerful discriminator between General Relativity (GR) and its extensions.

In modified gravity theories, deviations from GR alter the hydrostatic equilibrium condition through modifications of the Tolman--Oppenheimer--Volkoff (TOV) equations. As emphasized in recent studies, such deviations can either enhance or suppress the maximum mass depending on how the effective gravitational coupling and pressure support are modified at high densities. Importantly, these effects are non-perturbative and become significant only in the deep interior of neutron stars, where curvature and matter densities are extreme.

Within conservative $f(R,T)$ gravity, the explicit coupling between matter and geometry introduces additional terms in the equilibrium equations that effectively modify the balance between gravitational attraction and internal pressure. For the specific choice $h(T)=\lambda T$ adopted in this work, the coupling parameter $\lambda$ controls the strength of these deviations. Negative values of $\lambda$ generally act to weaken the effective gravitational pull inside the star, allowing configurations with larger masses for a given EOS, while positive values tend to reduce the maximum supported mass.

Our numerical results show that the maximum mass increases monotonically with decreasing $\lambda$ for all RMF-based EOSs considered, more described in the general conclusions reported in Refs.~\cite{Astashenok:2021btj, Astashenok:2021peo}, where it was demonstrated that modified gravity can mimic the effect of a stiffer EOS and thereby shift the maximum mass to higher values. However, we also find that this enhancement is EOS-dependent: softer EOSs benefit more significantly from modified gravity effects, whereas already stiff EOSs exhibit only modest changes.

A crucial point emphasized in the literature \cite{Astashenok:2021btj, Astashenok:2021peo} is the degeneracy between gravitational modifications and microphysical uncertainties in the EOS. Our results confirm this observation. In particular, certain RMF parameter sets that marginally fail to satisfy the $2\,M_\odot$ constraint in GR become compatible with observational bounds once moderate positive values of $\lambda$ are introduced. Conversely, excessively stiff EOSs that already predict overly large radii in GR cannot be reconciled with observational constraints even within the $f(R,T)$ framework, despite achieving higher maximum masses.

This highlights an important physical implication: modified gravity cannot arbitrarily rescue unrealistic EOSs. Instead, the combined constraints from maximum mass, radius measurements (e.g., NICER), and tidal deformability (from GW170817) jointly restrict both the EOS and the allowed range of the modified gravity parameter $\lambda$. Our analysis therefore supports the conclusion that neutron star maximum mass measurements serve as a complementary probe to gravitational-wave and X-ray observations in testing extended theories of gravity.

Overall, the behavior of the maximum mass in conservative $f(R,T)$ gravity observed in this work is fully consistent with earlier investigations of compact stars in modified gravity. At the same time, our results extend these studies by employing realistic RMF EOSs and by explicitly confronting the theoretical predictions with multi-messenger observational constraints, thereby strengthening the astrophysical viability of the model

\begin{table}[ht]
\centering
\begin{tabular}{lccccc}
\hline\hline
EOS & $\lambda$ & $M_{\max}\,(M_\odot)$ & $R_{\max}$ (km) & $R_{1.4}$ (km) & Observational status \\
\hline
DDH$_\delta$ & $0.4$ & $\sim 2.09$ & $\sim 11.06$ & $\sim 12.5$ & Allowed \\
DDH$_\delta$ & $0.6$ & $\sim 2.08$ & $\sim 11.00$ & $\sim 12.4$ & Allowed \\
TW            & $-0.4$ & $\sim 2.09$ & $\sim 10.70$  & $\sim 12.36$ & Allowed \\
TW            & $-0.2$ & $\sim 2.07$ & $\sim 10.60$ & $\sim 12.31$ & Allowed \\
DD2           & $0.8$ & $\sim 2.33$ & $\sim 11.72$ & $\sim 12.95$ & Marginal \\
NL3           & any    & $> 2.6$     & $> 13.1$   & $> 14.0$  & Disfavored \\
GM1           & any    & $> 2.3$     & $> 11.8$   & $> 13.4$  & Disfavored \\
TM1           & any    & $> 2.2$     & $> 12.1$     & $> 14.0$  & Disfavored \\
\hline\hline
\end{tabular}
\caption{Maximum gravitational mass $M_{\max}$ and corresponding radius $R_{\max}$, $R_{1.4}$ for different RMF equations of state in conservative $f(R,T)=R+\lambda T$ gravity. The results are compared with the qualitative bounds discussed in extended gravity studies such as \cite{Astashenok:2021btj}.}
\label{tab:maxmass_comparison}
\end{table}

Following the approach adopted in extended gravity studies, we analyze the maximum gravitational mass of neutron stars as a key diagnostic of both dense-matter physics and the underlying theory of gravity. In contrast to Ref.~\cite{Astashenok:2021btj}, where both baryonic and gravitational maximum masses were explicitly computed within $f(R)$ gravity, the present work focuses on the gravitational maximum mass obtained from the modified Tolman--Oppenheimer--Volkoff equations in 
conservative $f(R,T)$ gravity.

Our results demonstrate that the matter--geometry coupling parameter $\lambda$ systematically shifts the maximum mass configuration, but its effect remains strongly degenerate with the stiffness of the equation of state. Density-dependent RMF models such as DDH$_\delta$ and TW are able to simultaneously satisfy the $2\, M_\odot$ pulsar constraint, NICER radius bounds, and GW170817 tidal deformability limits for specific values of $\lambda$. In contrast, nonlinear RMF models (NL3, GM1, TM1), although capable of producing very large maximum masses, remain incompatible with radius and tidal constraints even within the modified gravity framework.

This behavior closely parallels the conclusions of \textit{Astashenok et. al}, where it was shown that extended gravity can moderately increase the maximum supported mass of neutron stars, but cannot compensate for unrealistically stiff equations of state. Our analysis, therefore, reinforces the view that maximum mass constraints provide a powerful and robust discriminator between viable dense-matter models and modified gravity 
effects in the strong-field regime.

An important question in the context of extended theories of gravity concerns the maximum mass that a neutron star can support against gravitational collapse. While modified gravity models can enhance the maximum mass relative to General Relativity (GR), recent studies have demonstrated that such enhancements are subject to robust theoretical limits. In particular, analyses in $f(R)$ gravity based on causal equations of state have shown that the maximum mass of static neutron stars remains bounded even under extreme assumptions on the stiffness of dense matter \cite{Astashenok:2021btj, Astashenok:2021peo}.

In GR, the maximum mass is determined by the turning-point criterion,
\begin{equation}
\left.\frac{dM}{d\rho_c}\right|_{\rho_c=\rho_c^{\max}} = 0 ,
\end{equation}
where $\rho_c$ is the central energy density. This condition continues to hold in modified gravity theories, although the location of the turning point and the corresponding maximum mass are shifted due to additional gravitational contributions. In conservative $f(R,T)$ gravity, the modified Tolman--Oppenheimer--Volkoff equations include an extra matter--geometry coupling term proportional to $\lambda$, which effectively alters the pressure gradient and the balance between gravitational attraction and internal pressure support.

The studies in Refs.~\cite{Astashenok:2021btj, Astashenok:2021peo} emphasize that, even in modified gravity, the maximum neutron star mass is constrained by fundamental physical requirements, most notably causality. Imposing the causal condition on the speed of sound,
\begin{equation}
c_s^2 = \frac{dP}{d\rho} \leq c^2 ,
\end{equation}
They demonstrate that the maximum gravitational mass of static neutron stars does not exceed $\sim 3\,M_{\odot}$, even for maximally stiff equations of state. This result holds for a wide class of $f(R)$ models and realistic matching to low-density nuclear equations of state, indicating that modified gravity can shift but not arbitrarily increase the upper mass limit.

Our results in conservative $f(R,T)$ gravity are fully consistent with this general picture. We find that increasing the matter--geometry coupling $\lambda$ can enhance the maximum mass for a given EOS by modifying the effective pressure support inside the star. However, this enhancement is limited and strongly dependent on the microphysical properties of dense matter. In particular, EOSs that fail to satisfy radius and tidal deformability constraints in GR remain disfavored even when modified gravity effects are included.

Furthermore, Refs.~\cite{Astashenok:2021btj, Astashenok:2021peo} highlight a strong degeneracy between gravitational modifications and the stiffness of the EOS: an increase in the modified gravity coupling can mimic the effect of a stiffer EOS, and vice versa. This degeneracy is also evident in our $f(R, T)$ analysis, where similar maximum masses can be obtained from different combinations of EOS parameters and $\lambda$. Consequently, neutron star maximum mass measurements alone are insufficient to disentangle modified gravity effects from dense-matter physics.

These findings reinforce the conclusion that modified gravity alone cannot resolve the neutron star mass-gap problem or compensate for unrealistic equations of state. Instead, robust constraints on extended gravity theories require the combined use of maximum mass measurements, radius determinations, tidal deformability from gravitational-wave observations, and physically consistent EOS modelling. In this sense, neutron stars continue to serve as powerful probes of gravity in the strong-field regime, while remaining subject to fundamental limits imposed by causality and microphysics.}

\subsection{Correlation between different physical observables of Neutron Stars}
In the present study, we present the $4 \times 4$ matrix for the correlation coefficient obtained for the modified gravity parameters $\lambda$  \& Neutron star observables in Figure (\ref{corel}) before applying astrophysical constraints using equations Pearson’s formula (left panel) and Kendall rank (right panel). The Kendall rank correlation can be considered a non-parametric test that allows us to quantify the strength of dependence between two variables. Unlike Pearson correlation, it accounts for non-linearity in the correlation, making it suitable for analyzing relationships that may not follow a linear pattern.

\begin{itemize}
    \item We see a strong correlation of $M_{max}$(r $= 0.85$) {\footnote{Here, r is the correlation coefficient}} with $R_{max}$ in the case of Pearson correlation where it boils down to $0.67$ for Kendall correlation. But it is negatively correlated with $\lambda$ ($-0.18$ and $-0.23$) for Pearson \& Kendall correlation respectively.
    \item For Pearson correlation   $\lambda$  is very weakly negatively correlated with $R_{1.4}$ and  $R_{max}$ (r $-0.16, -0.12$) respectively. But, the same is (r$ = -0.17,-0.2$) in the case of the Kendall correlation.
    \item $R_{1.4}$  has a high moderated correlation of $0.71$ with $M_{max}$ in Pearson formulas whereas the same with Kendal correlation give less moderated values of $0.63$ respectively.
    \item A very strong correlation of $R_{1.4}$ i.e (r = $0.96$) with $M_{max}$ is obtained from this matrix using the Pearson correlation formula, whereas Kendall correlation coefficient gives the moderate value of $0.95$ for the same.
\end{itemize}

\begin{widetext}
\begin{table}[ht]
    \centering
    \renewcommand{\arraystretch}{1.5} 
    \setlength{\tabcolsep}{12pt} 
    \captionsetup{width=\textwidth}
    \begin{tabular}{|c|c|c|c|c|c|c|}
        \hline
        $\lambda$ & DD$2$           & DDH$_\delta$          & TW            & NL$3$           & GM$1$          & TM$1$          \\ \hline
        -0.8      & 0.702207\textbf{237}   & 0.542589\textbf{312}  & 0.576042\textbf{752}   & 0.788966\textbf{869}   & 0.667026\textbf{997}  & 0.436861\textbf{370}  \\ 
        -0.6      & 0.702207\textbf{236}   & 0.542589\textbf{310}  & 0.576042\textbf{750}   & 0.788966\textbf{867}  & 0.667026\textbf{996}  & 0.436861\textbf{369}  \\ 
        -0.4      & 0.702207\textbf{235}   & 0.542589\textbf{308} & 0.576042\textbf{749}   & 0.788966\textbf{866}  & 0.667026\textbf{994}  & 0.436861\textbf{368}  \\ 
        -0.2      & 0.702207\textbf{233}   & 0.542589\textbf{307}  & 0.576042\textbf{747}   & 0.788966\textbf{865}   & 0.667026\textbf{993}  & 0.436861\textbf{366}  \\ 
         (0 $=$ GR)      & \underline{0.702207232}   & \underline{0.542589305}  & \underline{0.576042745}   & \underline{0.788966863}   & \underline{0.667026991}  & \underline{0.436861365}  \\ 
         0.2      & 0.702207\textbf{231}   & 0.542589\textbf{304}  & 0.576042\textbf{743}   & 0.788966\textbf{862}   & 0.667026\textbf{989}  & 0.436861\textbf{364}  \\ 
         0.4      & 0.702207\textbf{229}   & 0.542589\textbf{302}  & 0.576042\textbf{741}   & 0.788966\textbf{861}   & 0.667026\textbf{988}  & 0.436861\textbf{362}  \\ 
         0.6      & 0.702207\textbf{228}   & 0.542589\textbf{300}  & 0.576042\textbf{739}   & 0.788966\textbf{859}   & 0.667026\textbf{986}  & 0.436861\textbf{359}  \\ 
         0.8      & 0.702207\textbf{227}   & 0.542589\textbf{298}  & 0.576042\textbf{738}   & 0.788966\textbf{857}   & 0.667026\textbf{985}  & 0.436861\textbf{357}  \\ 
        \hline
    \end{tabular}
    \caption{Maximum $c_s^2/c^2$ values for taken EOS models for different values of $\lambda$, with $\lambda = 0$ returns to the GR value of the same}
    \label{tab:maximum-cs2}
\end{table}
\end{widetext}

\begin{widetext}

\begin{figure}[H]
\centering
\includegraphics[width=0.95\linewidth]{Figure6.png}\label{fig:corelation} 

\caption{\justifying  Correlation matrix for $\lambda$ with $M_{max}$, $R_{max}$,  and $R_{1.4}$ for all outputs of given EOS. The left panel uses Pearson’s formula, while the right uses the Kendall rank coefficient.}
\label{corel}
\end{figure}

\end{widetext}

\section{Conclusions}
\label{sec:concl}
{\color{black}
In this work, we have investigated the structure and observable properties of neutron stars within the framework of conservative $f(R, T)$ gravity, adopting the specific functional form $f(R, T)=R+\lambda T$, where $\lambda$ denotes the matter-geometry coupling parameter. Using a set of realistic relativistic mean-field (RMF) equations of state (EOSs) with diverse microphysical properties, many of which are constrained by experimental nuclear physics and many-body calculations, we solved the modified Tolman-Oppenheimer-Volkoff equations and confronted the resulting stellar configurations with multimessenger observational constraints from massive pulsars, NICER X-ray observations, and the binary neutron star merger GW170817.

To delineate the allowed regions of both dense-matter microphysics and modified gravity, we imposed the observational lower bound on the maximum neutron star mass, $M_{\max} \simeq 2.07\,M_{\odot}$, together with the GW170817-inferred radius constraint at canonical mass, $10.62~\mathrm{km} < R_{1.4} < 12.83~\mathrm{km}$. The EOSs considered span a broad range of stiffness, enabling a systematic exploration of the interplay between gravitational modifications and nuclear physics under joint electromagnetic and gravitational-wave constraints.

We analyzed two broad classes of RMF EOSs: density-dependent linear models (DD2, DDH$_\delta$, TW) and nonlinear interacting models with meson self-couplings (NL3, GM1, TM1). Our results demonstrate that density-dependent EOSs are significantly more compatible with the combined observational constraints within conservative $f(R, T)$ gravity. In particular, DDH$_\delta$ satisfies both GW170817 and NICER constraints for $\lambda=0.4$ and $0.6$, while the TW EOS remains viable for $\lambda=-0.4$ and $-0.2$. The DD2 EOS marginally overlaps the observational confidence regions but fails to simultaneously satisfy all constraints, yielding a maximum mass of $2.33\,M_{\odot}$ at $\lambda=0.8$. In contrast, the nonlinear EOSs NL3, GM1, and TM1, despite supporting large maximum masses, remain incompatible with the combined radius and tidal deformability bounds even after incorporating modified gravity effects.

A detailed examination of the mass-central density relation in the range $\rho_c \sim (2.5$--$4)\times10^{15}\,\mathrm{g\,cm^{-3}}$ reveals that, for lower values of $\lambda$, the maximum mass is achieved at comparatively lower central densities, beyond which the stellar mass decreases with increasing $\rho_c$, signaling the onset of instability. This behavior is observed across all EOSs and reflects the role of the matter-geometry coupling in shifting the stability boundary of compact configurations without altering its qualitative structure.

Our correlation analysis shows a strong positive correlation between the maximum mass and the corresponding radius, with Pearson and Kendall coefficients of $r=0.85$ and $0.67$, respectively. Conversely, the maximum mass exhibits a weak negative correlation with the coupling parameter $\lambda$, indicating that more negative values of $\lambda$ tend to favor effectively stiffer stellar configurations. We also verified that the modified pressure contributions preserve causality for all physically viable models, with the squared speed of sound satisfying $c_s^2/c^2<1$ throughout the stellar interior.

Importantly, our results align with recent studies of neutron stars in modified gravity \cite{Astashenok:2021btj, Astashenok:2021peo}, which demonstrate that even under extreme assumptions, such as maximally stiff, causal equations of state, the maximum mass of static neutron stars remains bounded at $\sim 3\,M_{\odot}$. In this context, our $f(R, T)$ analysis confirms that modified gravity can shift the maximum mass and mimic EOS stiffening, but cannot arbitrarily increase the upper mass limit. This highlights a strong degeneracy between dense-matter microphysics and gravitational modifications, implying that maximum mass measurements alone are insufficient to disentangle the two.

These findings reinforce the conclusion that modified gravity effects cannot compensate for unrealistic equations of state. Only EOSs that are already consistent with nuclear physics and multimessenger observational constraints remain viable within extended gravity frameworks. Consequently, robust tests of gravity in the strong-field regime require the combined use of maximum mass measurements, radius determinations, tidal deformability bounds, and physically consistent EOS modeling.

Finally, we emphasize that the conservation of the energy-momentum tensor is central to the physical viability of $f(R, T)$ gravity. By adopting the conservative functional form $f(R, T)=R+\lambda T$, we ensure $\nabla^{\mu}T_{\mu\nu}=0$, thereby avoiding unphysical particle creation or destruction in static equilibrium configurations \cite{dosSantos:2018nmu}. Our results demonstrate that conservative $f(R, T)$ gravity admits physically consistent neutron star solutions compatible with current observational constraints, reinforcing its relevance as a testable extension of General Relativity. In this broader context, our study supports the emerging view that the observed upper bounds on neutron star masses and their possible connection to the neutron star-black hole mass gap are governed by a nontrivial interplay between dense-matter microphysics and strong-field gravitational dynamics, rather than by modified gravity effects alone.}

\section{Acknowledgment}

P. Mahapatra would like to thank BITS Pilani K K Birla Goa campus for providing a good computational facility for this project and fellowship support. We thank Dr P Ajith (ICTS, Bangalore) for the useful discussions and insightful comments on this work. We thank Dr. Subhadip Sau (Jhargram Raj College, Jhargram, West Bengal) for some relevant ideas regarding the plot and comments on our EoS. Dr. Prasanta Kumar Das would like to thank Dr. Madhukar Mishra of BITS Pilani, Pilani campus for useful discussions related to Neutron Star.

\appendix
\section{Mass vs Central Energy density {$\rho_c$} plots}
\label{mass-rhoccurve}

\subsection{{$R + \lambda T$} model}

\begin{widetext}

\begin{figure}[H]
\centering
\subfigure[\  Mass$(M_\odot)$  vs $\rho_c$ ($10^{15}  g /cm^3$ ) ]{\includegraphics[width=0.49\linewidth]{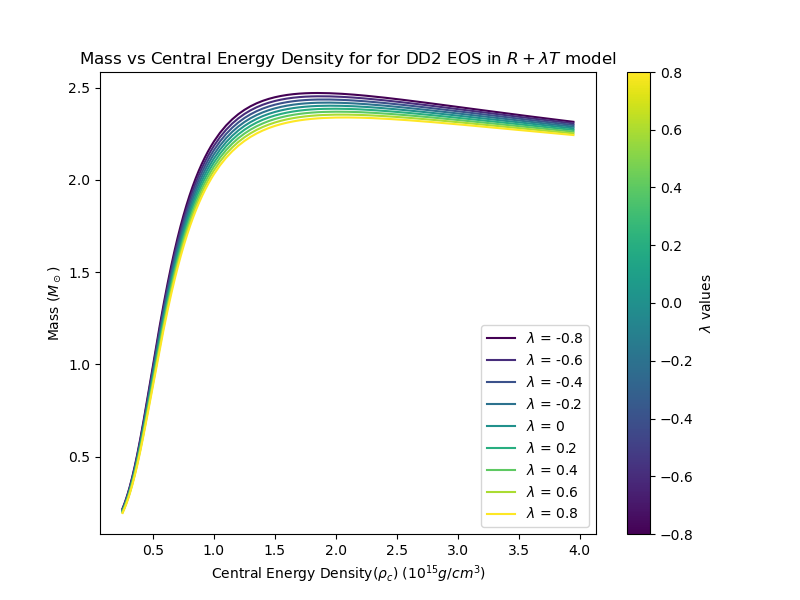}\label{fig:DD2_MVRhocnumberofstars}}
\subfigure[\  Mass$(M_\odot)$  vs $\rho_c$ ($10^{15}  g /cm^3$ ) ]{\includegraphics[width=0.49\linewidth]{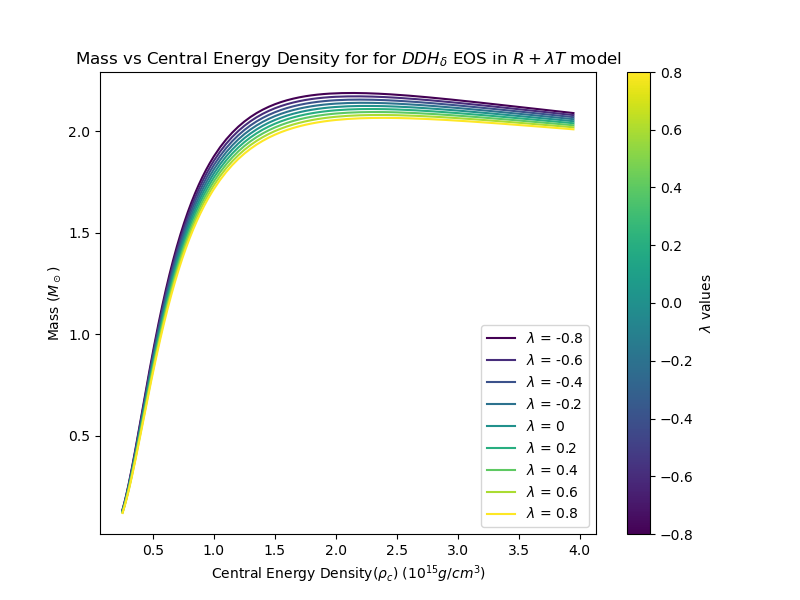}\label{fig:DDH_MVRhocnumberofstars}}
\subfigure[\  Mass$(M_\odot)$  vs $\rho_c$ ($10^{15}  g /cm^3$ ) ]{\includegraphics[width=0.49\linewidth]{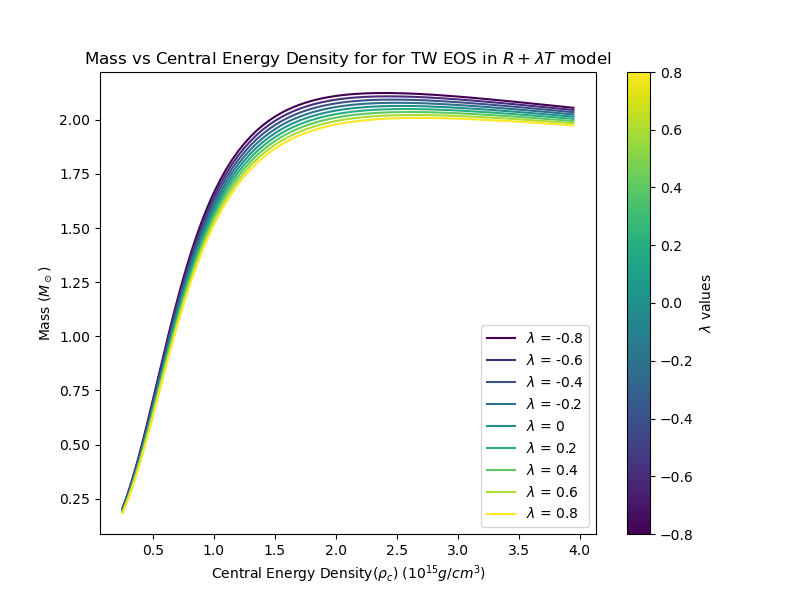}\label{fig:TW_MVRhocnumberofstars}}
\subfigure[\  Mass$(M_\odot)$  vs $\rho_c$ ($10^{15}  g /cm^3$ ) ]{\includegraphics[width=0.49\linewidth]{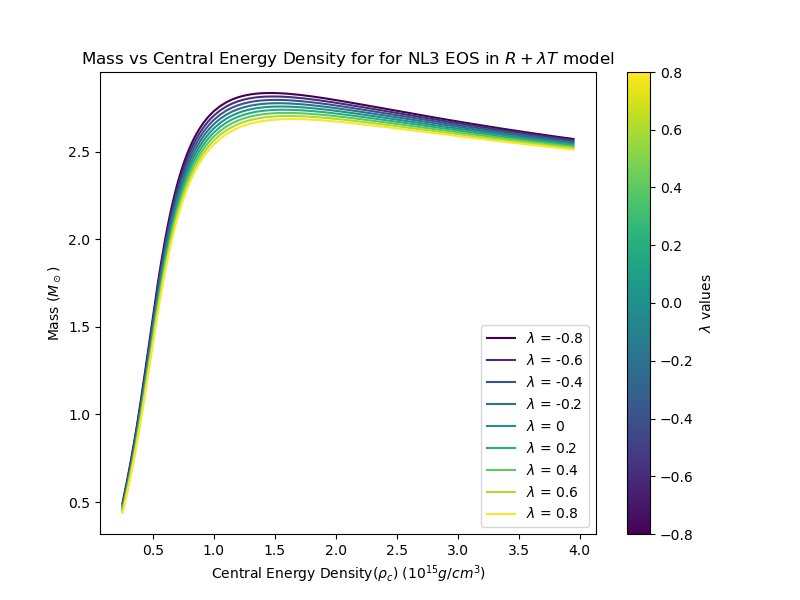}\label{fig:NL3_MVRhocnumberofstars}}
\subfigure[\  Mass$(M_\odot)$  vs $\rho_c$ ($10^{15}  g /cm^3$ ) ]{\includegraphics[width=0.49\linewidth]{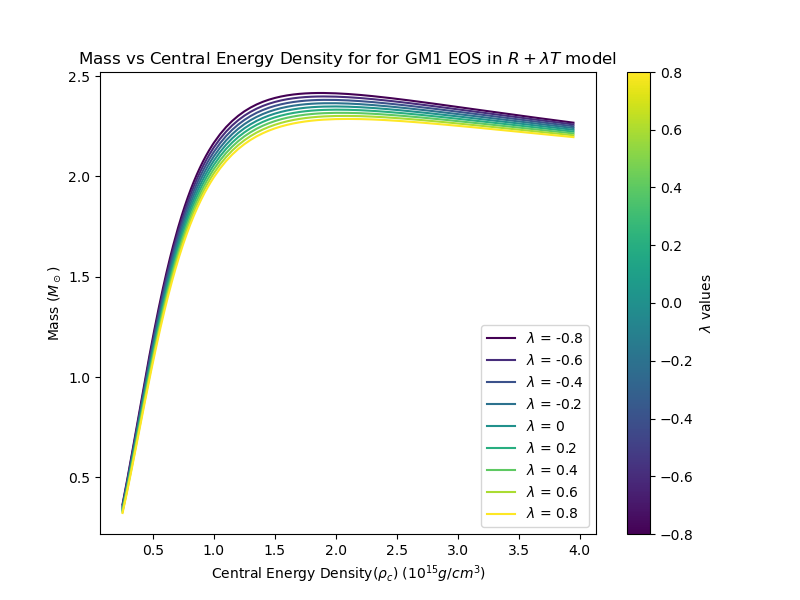}\label{fig:GM1_MVRhocnumberofstars}}
\subfigure[\  Mass$(M_\odot)$  vs $\rho_c$ ($10^{15}  g /cm^3$ ) ]{\includegraphics[width=0.49\linewidth]{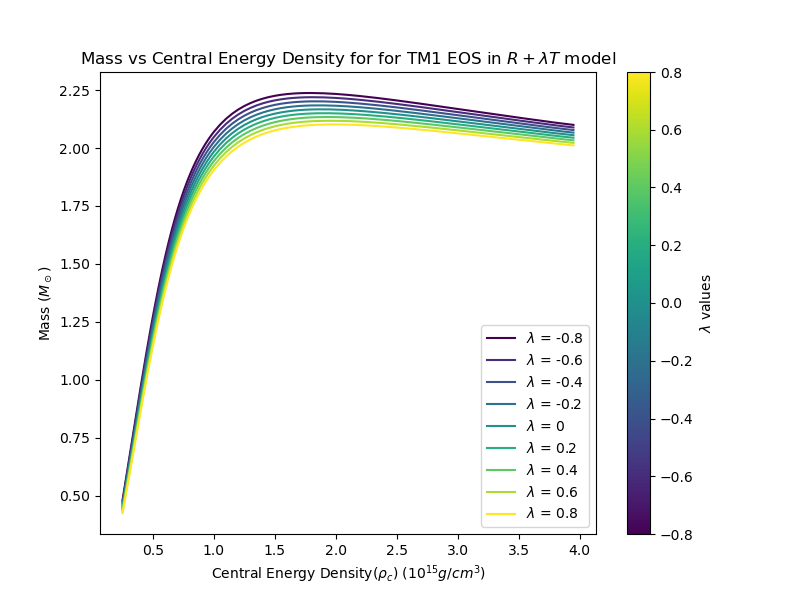}\label{fig:TM1_MVRhocnumberofstars}}
\caption{\justifying Mass($M_\odot$) vs central energy density $\rho_c$ ($10^{15}  g /cm^3$ )  for a number of Neutron Stars in the given range of $\rho_c$ for \textbf{DD2, DDH$_\delta$, TW, NL3, GM1, TM1 EOS}} 
\label{LAMBDAT_ALL-RHO}
\end{figure}

\end{widetext}

\bibliographystyle{./utphys1}
\bibliography{reference}
\end{document}